\documentclass[journal,12pt,onecolumn,draftclsnofoot]{IEEEtran}
\IEEEoverridecommandlockouts
\usepackage{cite}
\usepackage{amsmath,amssymb,amsfonts}
\usepackage{algorithmic}
\usepackage{graphicx,caption,subcaption}
\usepackage{textcomp}
\usepackage{xcolor,siunitx}
\usepackage[none]{hyphenat}
\usepackage[english,ruled,lined,linesnumbered]{algorithm2e} 
\def\BibTeX{{\rm B\kern-.05em{\sc i\kern-.025em b}\kern-.08em
		T\kern-.1667em\lower.7ex\hbox{E}\kern-.125emX}}
\newcommand{\B}[1]{\mathbf{#1}}

\newcommand{\R}[1]{\mathrm{#1}}
\newcommand{\bof}[1]{{\mbox{\boldmath$#1$}}}

\newcommand{\al}[1]{\textcolor{black}{{#1}}}
\newcommand{\gf}[1]{\textcolor{black}{{#1}}}
\newcommand{\fa}[1]{\textcolor{black}{{#1}}}

\newcommand{\nosemic}{\renewcommand{\@endalgocfline}{\relax}}
\newcommand{\dosemic}{\renewcommand{\@endalgocfline}{\algocf@endline}}
\let\oldnl\nl
\newcommand{\nonl}{\renewcommand{\nl}{\let\nl\oldnl}}
\makeatother

\begin{document}
	
	\title{Two-Dimensional Channel Parameter Estimation for IRS-Assisted Networks} 
	\author{Fazal-E-Asim,~\IEEEmembership{Senior~Member,~IEEE}, Andr\'e L. F. de Almeida,~\IEEEmembership{Senior~Member,~IEEE}, Bruno Sokal,~\IEEEmembership{Member,~IEEE},  Behrooz Makki,~\IEEEmembership{Senior~Member,~IEEE}, and G\'abor Fodor,~\IEEEmembership{Senior~Member,~IEEE}
 \thanks{Fazal-E-Asim, Bruno Sokal, and Andr\'e L. F. de Almeida all are with the Wireless Telecommunications Research Group (GTEL), Department
	of Teleinformatics Engineering, Federal University of Ceara, Brazil.
	Behrooz Makki is with Ericsson Research, Ericsson, 417 56 Gothenburg, Sweden.
	G\'abor Fodor is with Ericsson Research, 16480 Stockholm, and also with the Division of Decision and Control, KTH Royal Institute of Technology, 11428 Stockholm, Sweden. 
	e-mail: \{fazalasim, brunosokal, andre\}@gtel.ufc.br. \{behrooz.makki, gabor.fodor\}@ericsson.com.} }
%
	%

\maketitle

\begin{abstract}
	This paper proposes a pilot decoupling-based two-dimensional channel parameter estimation method for intelligent reflecting surface (IRS)-assisted networks. We exploit the combined effect of Terahertz sparse propagation and \gf{the} geometrical structure of arrays deployed at \gf{the base station, the IRS}, and the user equipment to develop a low-complexity channel parameter estimation method. By means of \gf{a} new pilot design along \gf{the} horizontal and vertical domains, the overall \al{channel parameter estimation} problem is decoupled into different domains. 
 Furthermore, with this decoupling, it is possible to simultaneously sense/estimate the channel parameters \gf{and} to communicate with the sensed node. \gf{Specifically}, we derive two estimators by decoupling the global problem into sub-problems and exploiting the built-in tensor structure of the \gf{sensing/estimation} problem by means of multiple rank-one approximations. The Cram\'er-Rao lower bound is derived to assess the performance of the proposed estimators. We show that the two proposed methods yield accurate parameter estimates and outperform state-of-the-art methods in terms of complexity. \al{The tradeoffs between performance and complexity offered by the proposed methods are discussed and numerically assessed.}  
\end{abstract}


\section{Introduction}

\IEEEPARstart{I}{ntelligent} reflecting surface (IRS) has been recently considered as a possible method for the enhancement of wireless communications in terms of spectral-efficiency (SE) and energy efficiency (EE) \cite{Huang_Zappone_2019,Basar_2019,Gong_2020,Guo_BM_2022,AlouiniBM_2022,Sokal_2022}. Moreover, utilizing millimeter-wave (mmWave) and Terahertz (THz) or (\SI{}{\mu\meter}Wave)  bands \gf{facilitates} a dense deployment of antennas at the base-station (BS) and the user equipment (UE) devices. Unfortunately, using high frequencies implies high path loss as well as \gf{high} penetration losses. To deal with this problem, IRS's having a large number of reflecting elements can be deployed to better focus the radiated energy towards the intended user/location. On the other hand, dealing with  large IRS panels in conjunction with massive \gf{multiple input multiple output (MIMO)} systems makes channel estimation a challenging task \cite{Boyu_Ning_2021}, especially for passive IRS structures, where the bulk of signal processing and complexity associated with the estimation of the involved large-scale channels is transferred to the end nodes of the network \cite{Renzo_2020}. Moreover, at high frequencies, the propagation channel is dominated by a few paths, which are more susceptible to blockages decreasing the overall coverage, throughput, and quality of service. In this scenario, IRS is a candidate solution to overcome these impairments by providing alternative communication paths with controlled propagation properties. 

The \gf{architecture proposed in} \cite{Masood_2022} uses inductive matrix completion followed by a classical root multiple signal classification (MUSIC) scheme to estimate the angles of arrival and departure in IRS-assisted systems. The authors of \cite{Wang_Peilan_2020} exploit the inherent sparsity of the mmWave channel to jointly design beamforming and estimate the cascaded BS-IRS-UE channel. A two-stage sparse recovery problem is proposed in \cite{Ardah_2021}, where the first stage delivers the directions of departure and arrival estimates, while in the second stage, the cascaded channel is estimated from the previously extracted angular parameters. The method proposed in \cite{Gilderlan_2021,Paulo_2022,Gilderlan_2022} relies on  tensor modeling to estimate the cascaded channel by alternating between the estimates of the individual channels.

Sequential channel estimation methods are proposed in \cite{Wang_Zhaorui_2020, MirzaJawad_2021, Swindlehurst_2022}, where the BS-UE and different direct or IRS-assisted channels are estimated sequentially.  The work \cite{ZihanYang_2022} considers a one-bit IRS for the direction of arrival (DoA) estimation by solving an atomic-norm-based sparse recovery problem, while \cite{BenediktJoham_2022} and \cite{ZhenKai_2022} propose neural network-based channel estimation methods for IRS-assisted networks. 

The authors of \cite{GuanXinrong_2022} propose a channel estimation technique based on anchor nodes by exploiting partial channel state information of the common BS-IRS link. The work in \cite{ChenXiao_2021} studies joint DoA and path gain estimation with limited radio-frequency chains and pilots, while \cite{ZhouPan_2022} proposes a cascaded channel estimation method to minimize the pilot overhead. The authors of \cite{JinZhang_2022} present a channel estimation strategy that exploits the low-rank structure of the involved channels using multiple residual dense networks. The work \cite{MirzaJawad_2021}  derives a two-stage channel estimation method, wherein the first stage is the direct \gf{MIMO}
channels between the end terminals are estimated, while in the second stage, the channel estimation problem is recast as a dictionary learning problem.

The authors in \cite{LiuDerrick_2022} model the BS-IRS-UE channel estimation as a denoising problem and then introduce deep residual learning to \gf{estimate the channel parameters}. The work \cite{Papazafeiropoulos_2022} takes residual hardware impairments into account using a linear minimum mean square error-based estimator followed by beamforming design. Assuming an orthogonal frequency division multiplexing system,  \cite{Zheng_2020} \gf{proposes} sequential-user channel estimation and training design. The work \cite{Lin_Tian_2022} formulates the channel estimation problem as $l_1$-norm regularized optimization having fixed-rank constraints. Building on various IRS models, channel estimation using compressed sensing is proposed in \cite{Ruan_2022} and \cite{Taha_2021}.  Cram\'er-Rao lower bound (CRLB) optimization is studied in \cite{Song_2022}, where the DoA parameters are estimated with the help of the non-line of sight (NLOS) echo reflected \textit{via} the IRS.

The paper \cite{ChenZhen_2022} exploits the sparsity of mmWave channels to solve the channel estimation problem in the mmWave scenario. The work \cite{JiangTao_2021} introduces implicit channel estimation by jointly optimizing the IRS phase shifts and the active beamforming from the received pilots with the additional \textit{a priori} information of the user locations. \cite{KimChoi_2022} proposes a dominant single-path approximation-based technique to extract the effective channel parameters. An orthogonal matching pursuit-based scheme is used to estimate the direction of departure (DoD) parameters. The work \cite{Emenonye_2022} investigates the multipath channel estimation problem from a Bayesian perspective. A sparse representation of the involved communication channels is also exploited in \cite{He_2020} to estimate the channels in a two-stage approach. \al{In \cite{fazaleasim2023tensorbased}, the full geometry of the BS-IRS-UE channels is exploited, which allows transforming the channel estimation problem into a single tensor approximation problem.} One practical example of IRS is given in \cite{yang2023vialess,wang2023wideband}.

In this paper, we study the problem of 
two-dimensional channel parameter estimation in THz IRS-assisted networks. We propose a strategy that efficiently exploits the geometrical structure of the involved communication channels. \al{More specifically,} the channel estimation problem is decoupled into sub-problems in the horizontal and vertical domains. The decoupling arises due to the independent design of the pilot matrices along each domain as a Kronecker product of the vertical and horizontal components. Such a pilot design decouples the overall problem into horizontal and vertical sub-problems. Thanks to such a decoupling, channel estimation is \al{translated into} to solving Kronecker factorization problems, which are recast as rank-one tensor approximation problems. In this regard, we propose two different estimators that split the global parameter estimation/sensing problem into independent and smaller  horizontal and vertical sub-problems. Also, we derive analytical CRLB expressions for linear estimation of the involved parameters, which serves as a reference for comparison with our proposed estimators. The contributions of this work are summarized as follows:
\begin{itemize}
\item We propose an independent pilot design along the horizontal and vertical domains by exploiting the geometrical structure of the antenna arrays at the BS and the UE as well as the array of reflecting elements at the IRS, where the ultimate pilot design is the Kronecker product of the vertical and the horizontal pilot matrices. Such a decoupling of pilots allows us to solve the associated Kronecker factorization problems as rank-one tensor approximation problems and, consequently, to decouple the global estimation problem into two smaller and low-complex horizontal and vertical sub-problems.
\item By exploiting the resulting low-rank tensor structure of the received pilots, we formulate two channel estimation methods: (1) the hybrid Kronecker factorization and multi-rank-one approximations (HKMR) method, which \al{directly} exploits the Kronecker-structured pilots and to solve \al{channel estimation} as horizontal and vertical sub-problems and (2) the two-stage higher-dimensional rank-one approximations (TSHDR) method, which also exploits the structure of pilots and the channel geometry to split the global problem into horizontal and vertical rank-one tensor approximation sub-problems.
\item We derive analytical CRLB expressions for the associated linear estimator, which serves as a reference for comparison with our proposed HKMR and TSHDR channel parameter estimation methods.
\item We compare our proposed methods with state-of-the-art solutions, including a least-squares (LS) method, the Khatri-Rao factorization (KRF) method proposed in \cite{Gilderlan_2021}, and the higher-dimensional rank-one
approximations (HDR) method \cite{fazaleasim2023tensorbased}. Also, we study the complexity of the proposed schemes.
Our results show that the proposed methods outperform their competitors in terms of \al{channel estimation accuracy and yield improved sum-rate-based performance}, especially in the lower signal-to-noise-ratio $(\R{SNR})$ regime.
Specifically, the proposed estimators outperform the competing LS and KRF methods \cite{Gilderlan_2021} which do not exploit the two-dimensional decoupling of the involved channel matrices, as well as the reference CRLB. The TSHDR method \al{provides more accurate channel parameter estimates} at the cost of higher complexity as compared to the HKMR method. 
\end{itemize}
\textit{\textbf{Notation:}} Scalars are denoted by lower-case italic letters $(a,b,\dots)$, vectors by bold lower-case italic letters $(\bof{a},\bof{b},\dots)$, matrices by bold upper-case italic letters $(\bof{A},\bof{B},\dots)$, tensors are defined by calligraphic upper-case letters $(\mathcal{A},\mathcal{B},\dots)$, $\bof{A}^\R{T}$,$\bof{A}^\R{\ast}$,$\bof{A}^\R{H}$ stand for transpose, conjugate and Hermitian of $\bof{A}$, respectively. The operator $\R{vec}\left\{.\right\}$ vectorizes an $I \times J$ matrix argument, while $\R{unvec}_{(I \times J)}\left\{.\right\}$ does the opposite operation. Also, $\R{tr}\{.\}$ represents trace of the matrix. The operators $\otimes$, $\diamond$, $\circ$, and $\odot$ define the Kronecker, the Khatri-Rao, the outer product, and the Hadamard product, respectively. 
\al{The $n$-mode product of a tensor $\mathcal{X} \in \mathbb{C}^{I \times J \times K}$ and a matrix $\bof{A} \in \mathbb{C}^{I \times R}$ is denoted as $\mathcal{Y}=\mathcal{X}\times_n\bof{A}$, $n=1,2,3$.}
Also, $\R{diag}\{\B{a}\}$ represents the square diagonal matrix with elements of vector $\B{a}$ across the main diagonal. Finally, $\mathbb{E}[\cdot]$ is the expectation operator, and $\bof{I}$ is the identity matrix.

\begin{figure}[!t]
\centerline{\includegraphics[width=0.5\linewidth]{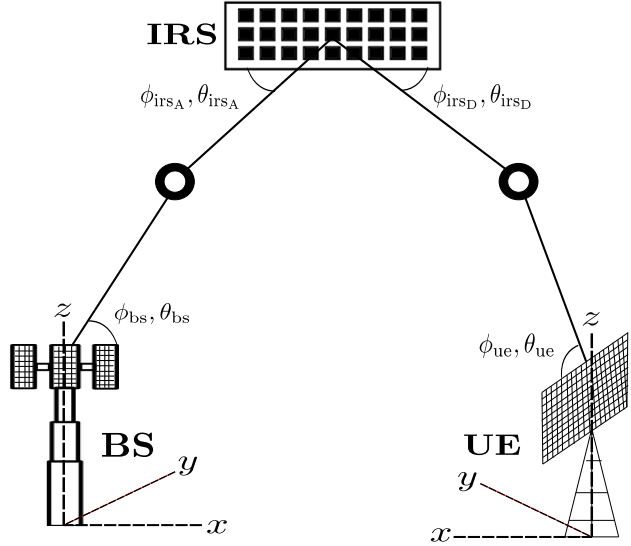}}
\caption{Illustration of an IRS-assisted MIMO communication system with two-dimensional parameter estimation.}
\label{fig:1}
\end{figure}

\section{System and Channel Model}
Consider a BS having $M$ antenna elements and communicating with a UE having $Q$ antenna elements \textit{via} an IRS having $N$ reflecting elements as shown in \figurename{~\ref{fig:1}}. A uniform rectangular array (URA) is deployed at both the BS and the UE.
The channel $\bof{H} \in \mathbb{C}^{N \times M}$ between the BS and the IRS, and the channel $\bof{G} \in \mathbb{C}^{Q \times N}$ between the IRS and the UE are represented by their respective geometrical models. Motivated by the THz communication characteristics and because in practice the IRSs will be deployed with proper network planning, we assume that the channel linking the BS to the IRS is dominated by its line-of-sight (LOS) path, i.e., the non-LOS (NLOS) terms are weak and assumed to be negligible. Therefore, $\bof{H}$ can be written as 
\begin{align}
\bof{H} =\bof{b}(\phi_{\text{irs}_\text{A}},\theta_{\text{irs}_\text{A}}) \bof{a}^\R{T}(\phi_{\text{bs}},\theta_{\text{bs}}), \label{H_sph} 
\end{align}
where $\bof{a}(\phi_{\text{bs}},\theta_{\text{bs}}) \in \mathbb{C}^{M \times 1}$ is the two-dimensional  steering vector at the BS, $\phi_{\text{bs}}$ is the azimuth of departure (AoD) and $\theta_{\text{bs}}$ as elevation of departure (EoD). Similarly, $\bof{b}(\phi_{\text{irs}_\text{A}},\theta_{\text{irs}_\text{A}}) \in \mathbb{C}^{N \times 1}$ is the two-dimensional IRS steering vector with  $\phi_{\text{irs}_\text{A}}$ being the azimuth of arrival (AoA) and $\theta_{\text{irs}_\text{A}}$ the elevation of arrival (EoA).
Similarly, assuming a LOS path between the IRS and the UE, the channel $\bof{G}$ is given as
\begin{align}
\bof{G} =\bof{q}(\phi_{\text{ue}},\theta_{\text{ue}}) \bof{p}^\R{T}(\phi_{\text{irs}_\text{D}},\theta_{\text{irs}_{\text{D}}}), \label{G_sph}
\end{align}
where $\bof{p}(\phi_{\text{irs}_\text{D}},\theta_{\text{irs}_{\text{D}}}) \in \mathbb{C}^{N \times 1}$ is the two-dimensional steering vector at the IRS, where $\phi_{\text{irs}_\text{D}}$ and $\theta_{\text{irs}_{\text{D}}}$ are the corresponding AoD and EoD, respectively. Similarly, $\bof{q}(\phi_{\text{ue}},\theta_{\text{ue}}) \in \mathbb{C}^{Q \times 1}$ is the two-dimensional channel steering vector at the UE, with $\phi_{\text{ue}}$ and $\theta_{\text{ue}}$ being the associated directional parameters. Without loss of generality, we can further assume that the channel attenuation coefficient has been absorbed by one of the steering vectors in \eqref{H_sph} and \eqref{G_sph}, which will not affect the channel parameter estimation methods.

\subsection{Kronecker Product-Based Channel Factorizations}
\gf{Consider} the BS array response in the $y\text{-}z$ plane, which can be modeled as
\begin{equation}
[\bof{a}(\phi_{\text{bs}},\theta_{\text{bs}})]_m = e^{-j\pi[(m_y-1)\sin\theta_{\text{bs}}\sin\phi_{\text{bs}} + (m_z-1)\cos\theta_{\text{bs}}]}, \label{ch_3_y-z}
\end{equation}
with $m = m_z + (m_y-1)M_z$, $m_y \in \{1,\dots,M_y\}$, and  $m_z \in \{1,\dots,M_z\}$, $M=M_yM_z$, where $M_y$ is the number of antenna elements along $y$-axis, and $M_z$ is the number of antenna elements along $z$-axis, as explained in \cite{Asim_2021}. Defining the spatial frequencies as $\mu_{\text{bs}} = \pi\sin\theta_{\text{bs}}\sin\phi_{\text{bs}}$ and $\psi_{\text{bs}} = \pi\cos\theta_{\text{bs}}$, the steering vector can be expressed as the Kronecker product between the two steering vectors as follows

\begin{equation}
\bof{a}(\mu_{\text{bs}},\psi_{\text{bs}}) = \bof{a}_y(\mu_{\text{bs}}) \otimes \bof{a}_z(\psi_{\text{bs}}) \in \mathbb{C}^{M \times 1}. \label{ch_3_kron_channel}
\end{equation}
where
\begin{eqnarray}
&&\bof{a}_y(\mu_{\text{bs}}) = \left[1, e^{-j\mu_{\text{bs}}} ,\dots, e^{-j(M_y-1)\mu_{\text{bs}}}\right]^\R{T} \in \mathbb{C}^{M_y \times 1},\nonumber \\
&&\bof{a}_z(\psi_{\text{bs}}) = \left[1, e^{-j\psi_{\text{bs}}} ,\dots, e^{-j(M_z-1)\psi_{\text{bs}}}\right]^\R{T} \in \mathbb{C}^{M_z \times 1}.\nonumber
\end{eqnarray}
Similarly, $\bof{b}(\phi_{\text{irs}_\text{A}},\theta_{\text{irs}_\text{A}})$ steering vector can be factorized as a Kronecker product of $\bof{b}_y(\mu_{\text{irs}_\text{A}})$ and $\bof{b}_z(\psi_{\text{irs}_\text{A}})$.  
Hence, $\bof{H}$ given in \eqref{H_sph} and $\bof{G}$ given in \eqref{G_sph} can be rewritten in terms of spatial frequencies and Kronecker products as
\begin{align}
\bof{H} &= \left[\bof{b}_y(\mu_{\text{irs}_\text{A}}) \otimes \bof{b}_z(\psi_{\text{irs}_\text{A}}) \right] \left[\bof{a}_y(\mu_{\text{bs}}) \otimes \bof{a}_z(\psi_{\text{bs}})\right]^\R{
	T} \label{H_spat}\\
\bof{G} &= \left[\bof{q}_y(\mu_{\text{ue}}) \otimes \bof{q}_z(\psi_{\text{ue}}) \right] \left[\bof{p}_y(\mu_{\text{irs}_\text{D}}) \otimes \bof{p}_z(\psi_{\text{irs}_\text{D}})\right]^\R{
	T}. \label{G_spat}
\end{align}

Using the Kronecker \al{product} properties $\left(\bof{A} \otimes \bof{B}\right)^\R{T} = \bof{A}^\R{T} \otimes \bof{B}^\R{T}$, and $\left(\bof{A} \otimes \bof{B}\right) \left(\bof{C} \otimes \bof{D}\right) = \left(\bof{AC} \otimes \bof{BD}\right)$, the BS-IRS channel $\bof{H}$ in \eqref{H_spat} can be rewritten as
\begin{align}
\bof{H} = \underbrace{\left[\bof{b}_y(\mu_{\text{irs}_\text{A}})   \bof{a}_y^\R{
		T}(\mu_{\text{bs}})\right]}_{\bof{H}_y} \otimes \underbrace{\left[\bof{b}_z(\psi_{\text{irs}_\text{A}}) \bof{a}_z^\R{
		T}(\psi_{\text{bs}})\right]}_{\bof{H}_z}, \label{final_H}
\end{align}
where $\bof{H}_y \in \mathbb{C}^{N_y \times M_y}$ is the channel along the horizontal (azimuth) domain and $\bof{H}_z \in \mathbb{C}^{N_z \times M_z}$ is the channel along the vertical (elevation) domain. Similarly, the IRS-UE channel $\bof{G}$ in \eqref{G_spat} can be factorized as
\begin{align}
\bof{G} = \underbrace{\left[\bof{q}_y(\mu_{\text{ue}})\bof{p}_y^\R{
		T}(\mu_{\text{irs}_\text{D}})   \right]}_{\bof{G}_y} \otimes \underbrace{\left[\bof{q}_z(\psi_{\text{ue}}) \bof{p}_z^\R{
		T}(\psi_{\text{irs}_\text{D}})\right]}_{\bof{G}_z},
\label{final_G}
\end{align}
where $\bof{G}_y \in \mathbb{C}^{Q_y \times N_y}$ is the channel along the horizontal (azimuth) domain and $\bof{G}_z \in \mathbb{C}^{Q_z \times N_z}$ is the channel along the vertical (elevation) domain.

\subsection{Signal Model}
To exploit the geometrical \al{channel} structure and to decouple the global \al{channel estimation} into smaller horizontal and vertical sub-problems, we design independent pilot matrices along each domain, such that the resulting pilot matrix is \gf{factorized as the Kronecker product of a horizontal and a vertical pilot matrix}, as follows
\begin{align}
\label{eq:S}
\bof{S} = \bof{S}_y \otimes \bof{S}_z \in \mathbb{C}^{M \times T},
\end{align}
where $\bof{S}_y \in \mathbb{C}^{M_y \times T_y}$ and $\bof{S}_z \in \mathbb{C}^{M_z \times T_z}$ are the pilot symbol matrices associated with horizontal and vertical domains, respectively, while $T = T_yT_z$ is the length of the pilot sequence, $T_y$ is the length of the horizontal pilot sequence, and $T_z$ is the length of the vertical pilot sequence. We assume that the horizontal and vertical pilot matrices are designed as Hadamard matrices of dimensions $M_y$ and $M_z$, respectively, with $M=M_yM_z$, for two-dimensional channel probing and sensing of the unknown parameters. Note that the resulting pilot symbol matrix is also a $M_yM_z$-dimensional Hadamard matrix. The motivation behind the use of Hadamard design is that it can be effectively modulated as  binary phase shift keying (BPSK) symbols, which helps to drive power amplifiers close to the saturation region, due to its constant modulus property. 
The received pilot signal at the UE \textit{via} the IRS by placing the $k$-th phase-shift pattern at the IRS is given by
\begin{equation}
\bof{X}_{k} = \bof{G} \R{diag}\left\{\bof{\omega}_{k}\right\}\bof{H}\bof{S} + \bof{V}_k \in \mathbb{C}^{Q \times T},
\label{recived_eq}
\end{equation}
\fa{where $\bof{V}_k \sim \mathcal{CN}\left(\bof{0}_{Q\times T},\sigma_n^2\bof{I}_{Q \times T}\right)$ is the circularly symmetric additive white Gaussian noise matrix with variance $\sigma_n^2$ and $\bof{R}_k$ is the noise covariance matrix assuming spatially and temporally uncorrelated noise as with $\bof{R}_k = \R{E}\left[\R{vec}\{\bof{V}_k\} \; \R{vec}\{\bof{V}_k\}^{\R{H}} \right]=\sigma_n^2 \bof{I}_{QT \times QT}$. The vector $\bof{\omega}_k$ holds the reflection  amplitude  and phase coefficients of the IRS used at the $k$-th block. Following the same principle adopted for the design of the pilot structure, we assume that $\bof{\omega}_k =\bof{\omega}_y^{(k)} \otimes \bof{\omega}_z^{(k)}$, where $\bof{\omega}_y^{(k)} \in \mathbb{C}^{N_y \times 1}$ and $\bof{\omega}_z^{(k)}\in \mathbb{C}^{N_z \times 1}$ are $y$ and $z$ domains IRS phase shift vectors used at the $k$-th block. We choose $\bof{\omega}_y^{(k)}$ and $\bof{\omega}_y^{(k)}$ as the $k$-th columns of matrices $\bof{\Omega}_y \in \mathbb{C}^{N_y \times K}$ and $\bof{\Omega}_z \in \mathbb{C}^{N_z \times K}$, respectively, 
which are constructed as
\begin{align}
\bof{\Omega}_y = \bof{W}_y \bof{\Psi},\quad \bof{\Omega}_z = \bof{W}_z \bof{\Phi}  \label{eq:omegayomegaz}
\end{align}
where $\bof{W}_y \in \mathbb{C}^{N_y \times K_y}$ and $\bof{W}_z \in \mathbb{C}^{N_z \times K_z}$ are (possibly truncated) Discrete Fourier Transform (DFT) matrices of size $K_y$ and $K_z$, respectively, while $\bof{\Psi} = \bof{I}_{K_y} \otimes \bof{1}_{K_z}^\R{T} \in \mathbb{R}^{K_y \times K}$ and $\bof{\Phi} = \bof{1}_{K_y}^\R{T} \otimes \bof{I}_{K_z} \in \mathbb{R}^{K_z \times K}$, with $K=K_yK_z$. This design implies that 
$\bof{\Omega}=[\bof{\omega}_1, \ldots, \bof{\omega}_K]=\bof{\Omega}_y \diamond \bof{\Omega}_z \in \mathbb{C}^{N \times K}$. Using (\ref{eq:omegayomegaz}), we have $\bof{\Omega}= \left(\bof{W}_y\bof{\Psi}\right) \diamond \left(\bof{W}_z\bof{\Phi}\right)=\left(\bof{W}_y \otimes \bof{W}_z\right)\left(\bof{\Psi} \diamond \bof{\Phi}\right)=\bof{W}_y \otimes \bof{W}_z$, since $\bof{\Psi} \diamond \bof{\Phi} = \bof{I}_K$. Hence, each column $\bof{\omega}_k$ of $\bof{\Omega}$ can be generated from the corresponding columns of the DFT codebooks $\bof{W}_y$ and $\bof{W}_z$ \textit{via} the Kronecker product. This structure was also adopted in \cite{Asim_2021,Fazal-E-Asim_2022} showing to have lower implementation complexity as well as higher energy efficiency. Moreover, assuming that $\bof{W}_y$ and $\bof{W}_z$ are row-orthonormal DFT matrices with $\bof{W}_y\bof{W}^\R{H}_y=\bof{I}_{N_y}$ and $\bof{W}_z\bof{W}^\R{H}_z=\bof{I}_{N_z}$, we have $\bof{\Omega}\bof{\Omega}^\R{H}=\bof{I}_N$.
}

\al{Using the proposed Kronecker-structured pilot and IRS phase shift structures, we can} rewrite \eqref{recived_eq} as
\begin{align}
\bof{X}_{k} = \left( \bof{G}_y \otimes \bof{G}_z\right) \left(\R{diag}\{\bof{\omega}_y^{(k)}\}\otimes \R{diag}\{\bof{\omega}_z^{(k)}\}\right) \left(\bof{H}_y \otimes \bof{H}_z\right) \left(\bof{S}_y \otimes \bof{S}_z\right) + \bof{V}_k.
\end{align}
Using property $\left(\bof{A} \otimes \bof{B}\right) \left(\bof{C} \otimes \bof{D}\right) = \left(\bof{AC}\right) \otimes \left(\bof{BD}\right)$ yields
\begin{align}
	\bof{X}_k =\underbrace{\left(\bof{G}_y\R{diag}\{\bof{\omega}_y^{(k)}\}\bof{H}_y\bof{S}_y\right)
}_{\bof{X}_y^{(k)}} \otimes \underbrace{\left(\bof{G}_z\R{diag}\{\bof{\omega}_z^{(k)}\}\bof{H}_z\bof{S}_z\right)}_{\bof{X}_z^{(k)}}  + \bof{V}_k,
\vspace{-2ex}
\end{align}
which can be further written in a compact form as
\begin{align}
\bof{X}_k = \bof{X}_y^{(k)} \otimes \bof{X}_z^{(k)} + \bof{V}_k,
\label{eq:kronpilotsX}
\end{align}
where $\bof{X}_y^{(k)} \in \mathbb{C}^{Q_y \times T_y}$ and $\bof{X}_z^{(k)} \in \mathbb{C}^{Q_z \times T_z}$ are  the $k$-th horizontal and vertical received pilot signals, respectively.
\al{Defining $\bof{U}^{(k)}_y=\bof{G}_y\R{diag}\{\bof{\omega}_y^{(k)}\}\bof{H}_y \in \mathbb{C}^{Q_y \times M_y}$ and $\bof{U}^{(k)}_z=\bof{G}_z\R{diag}\{\bof{\omega}_z^{(k)}\}\bof{H}_z \in \mathbb{C}^{Q_z \times M_z}$ as the $y$-domain and $z$-domain cascaded MIMO channels associated with the $k$-th training block, we have $\bof{X}_y^{(k)}=\bof{U}^{(k)}_y\bof{S}_y$ and $\bof{X}_z^{(k)}=\bof{U}^{(k)}_z\bof{S}_z$.}

\section{Estimation of Channel Parameters}
In this section, two estimators are derived to solve the channel parameter estimation problem, \al{which effectively exploits the explicit channel structure and the two-dimensional decoupling of the received pilots. The first estimator separately obtains the $y$-domain and $z$-domain BS and UE channel parameters in the first stage by solving rank-one tensor approximation problems, while extracting all the IRS-related channel parameters in the second stage \textit{via} rank-one matrix approximation. The second estimator simultaneously extracts BS, UE, and IRS channel parameters by solving  horizontal and vertical rank-one tensor approximation problems.} In the following, each proposed estimator is formulated.

\subsection{Hybrid Kronecker Factorization and Multi-rank-one Approximations (HKMR) Method}
\al{As shown in (\ref{eq:kronpilotsX}), the useful signal part of the received pilot signals has a Kronecker-product structure. The first step of the HKMR method consists of estimating the $y$-domain and $z$-domain received pilots from the overall received pilots. This problem is translated into minimizing the following cost function in the least squares sense}
\begin{equation}
\left\{\hat{\bof{X}}_y^{(k)},\hat{\bof{X}}_z^{(k)}\right\} = \arg\min_{\bof{}} \left \|\bof{X}_k - \bof{X}_y^{(k)} \otimes \bof{X}_z^{(k)}  \right\|_\R{F}^2, \label{LS_kron_probel}
\end{equation}
which can be recast as a rank-one matrix approximation \cite{Loan1992ApproximationWK}
\begin{align}
\left\{\hat{\bof{X}}_y^{(k)},\hat{\bof{X}}_z^{(k)}\right\}\!=\! \arg\min \left \| \bof{X}^\prime_k  \! -\! \bof{x}^{(k)}_z \bof{x}^{(k)\R{T}}_y  \right\|_\R{F}^2, \label{LS_Kron_prob_A}
\end{align}
\al{where $\bof{x}^{(k)}_y=\R{vec}\{\bof{X}_y^{(k)}\} \in \mathbb{C}^{Q_yT_y \times 1}$ and $\bof{x}^{(k)}_z=\R{vec}\{\bof{X}_z^{(k)}\} \in \mathbb{C}^{Q_zT_z \times 1}$, and $\bof{X}^\prime_k \in \mathbb{C}^{Q_zT_z \times Q_yT_y}$ is obtained by properly rearranging the entries of $\bof{X}_k$ (see \cite{Loan1992ApproximationWK} for details). The problem (\ref{LS_Kron_prob_A}) is simple and can be solved by any state-of-the-art rank-one matrix approximation procedure for which efficient solutions exist in the literature}. \al{Note that the estimates of the horizontal and vertical pilots involve a scaling ambiguity, i.e. $\hat{\bof{X}}_y^{(k)}=\alpha_k\bof{X}_y^{(k)}$ and $\hat{\bof{X}}_z^{(k)}=\beta_k\bof{X}_z^{(k)}$, where $\alpha_k$ and $\beta_k$ are the corresponding scaling factors that compensate each other, such that $\alpha_k\beta_k=1$.}

After estimating $\hat{\bof{X}}_y^{(k)}$ and $\hat{\bof{X}}_z^{(k)}$, independent matched filtering (MF) is applied along horizontal and vertical domains \al{using the known $y$-domain and $z$-domain pilot signals}  to estimate the corresponding \al{cascaded} channel matrices as
\begin{align}
&\al{\hat{\bof{U}}_y^{(k)}} = \hat{\bof{X}}_y^{(k)}\bof{S}_y^\R{H} \in \mathbb{C}^{Q_y \times M_y} \label{MF_y}\\
&\al{\hat{\bof{U}}_z^{(k)}} = \hat{\bof{X}}_z^{(k)}\bof{S}_z^\R{H} \in \mathbb{C}^{Q_z \times M_z}. \label{MF_z}
\end{align}
\al{Defining $\hat{\bof{u}}_y^{(k)} = \R{vec}\big\{\hat{\bof{U}}_y^{(k)}\big\} \in \mathbb{C}^{Q_yM_y \times 1}$ and $\hat{\bof{u}}_z^{(k)} = \R{vec}\big\{\hat{\bof{U}}_z^{(k)}\big\} \in \mathbb{C}^{Q_zM_z \times 1}$, and taking into account the scaling ambiguities, 
we have}
\begin{align}
&\hat{\bof{u}}_y^{(k)}= \left(\bof{H}_y^\R{T} \diamond \bof{G}_y\right)\bof{\omega}_y^{(k)}\alpha_k, \label{x_hathat_k_y}\\
&\hat{\bof{u}}_z^{(k)} = \left(\bof{H}_z^\R{T} \diamond \bof{G}_z\right)\bof{\omega}_z^{(k)}\beta_k.\label{x_hathat_k_z}
\end{align}
\al{Defining the matrices $\hat{\bof{U}}_y = [\hat{\bof{u}}_y^{(1)},\dots,\hat{\bof{u}}_y^{(K)}] \in \mathbb{C}^{Q_yM_y \times K}$ and $\hat{\bof{U}}_z = [\hat{\bof{u}}_z^{(1)},\dots,\hat{\bof{u}}_z^{(K)}] \in \mathbb{C}^{Q_zM_z \times K}$ that collect these estimates for $K$ blocks, we obtain}
\begin{align}
&\al{\hat{\bof{U}}_y} =  \left(\bof{H}_y^\R{T} \diamond \bof{G}_y\right)\bof{\Omega}_y \bof{D}(\bof{\alpha}) \in \mathbb{C}^{Q_yM_y \times K} \label{X_tilde_est_y}\\
&\al{\hat{\bof{U}}_z} =  \left(\bof{H}_z^\R{T} \diamond \bof{G}_z\right)\bof{\Omega}_z\bof{D}(\bof{\beta}) \in \mathbb{C}^{Q_zM_z \times K}\label{X_tilde_est_z}
\end{align} 
where $\bof{D}(\bof{\alpha}) = \R{diag}\left\{\bof{\alpha}\right\}, \bof{\alpha} = \left[\alpha_1,\dots,\alpha_K\right]^\R{T}$ and $\bof{D}(\bof{\beta}) = \R{diag}\left\{\bof{\beta}\right\}, \bof{\beta} = \left[\beta_1,\dots,\beta_K\right]^\R{T}$. \al{Since the scaling factors cancel each other, we have $\bof{D}(\bof{\alpha})\bof{D}(\bof{\beta})=\bof{I}_K$. 
From the rank-one structures of $\bof{H}_y$ and $\bof{G}_y$ defined in \eqref{final_H} and \eqref{final_G} respectively,  we  can expand \eqref{X_tilde_est_y} as
\begin{align}
\hat{\bof{U}}_y &= 
\Big(\big[\bof{a}_y(\mu_{\text{bs}})\bof{b}_y^\R{T}(\mu_{\text{irs}_\text{A}})\big] \diamond \big[ \bof{q}_y(\mu_{\text{ue}})\bof{p}_y^\R{T}(\mu_{\text{irs}_\text{D}})\big]\Big)\bof{\Omega}_y \bof{D}(\bof{\alpha}). \nonumber 
\end{align}
Applying the mixed property rule $\left(\bof{A} \otimes \bof{B}\right)\left(\bof{C}\diamond\bof{D}\right) = \left(\bof{AC}\right)\diamond \left(\bof{BD}\right)$, and  noting that $\bof{a}^\R{T} \diamond \bof{b}^\R{T} = \bof{a}^\R{T} \odot \bof{b}^\R{T}$, yields}
\begin{align}
\al{\hat{\bof{U}}_y}\!=\!\left(\bof{a}_y(\mu_{\text{bs}}) \otimes  \bof{q}_y(\mu_{\text{ue}})\right)\left( \bof{b}_y(\mu_{\text{irs}_\text{A}})\odot\bof{p}_y(\mu_{\text{irs}_\text{D}})\right)^\R{T}\bof{\Omega}_y \bof{D}(\bof{\alpha}),
\nonumber 
\end{align}
which can be written in a compact form as
\begin{align}
\al{\hat{\bof{U}}_y} = \left(\bof{a}_y(\mu_{\text{bs}}) \otimes  \bof{q}_y(\mu_{\text{ue}})\right)\bof{n}_y^\R{T}(\mu_{\text{y}})\bof{\Omega}_y \bof{D}(\bof{\alpha}),
\label{X_y}
\end{align}
\al{where} $\bof{n}_y(\mu_{\text{y}}) = \bof{b}_y(\mu_{\text{irs}_\text{A}})\odot\bof{p}_y(\mu_{\text{irs}_\text{D}})\in\mathbb{C}^{Ny \times 1}$ is the effective $y$-domain IRS steering vector, while $\mu_{\text{y}} = \mu_{\text{irs}_\text{A}}+ \mu_{\text{irs}_\text{D}} $ is the combined $y$th spatial frequency. \al{Analogously, by following the same steps, we can write $\hat{\bof{U}}_z$ as}
\begin{align}
\al{\hat{\bof{U}}_z} = \left(\bof{a}_z(\psi_{\text{bs}}) \otimes  \bof{q}_z(\psi_{\text{ue}})\right)\bof{n}_z^\R{T}(\psi_{\text{z}})\bof{\Omega}_z \bof{D}(\bof{\beta}), \label{X_z}
\end{align}
where $\bof{n}_z(\psi_{\text{z}}) = \bof{b}_z(\psi_{\text{irs}_\text{A}})\odot\bof{p}_z(\psi_{\text{irs}_\text{D}}) \in\mathbb{C}^{Nz \times 1}$ and $\psi_{\text{z}} = \psi_{\text{irs}_\text{A}}+ \psi_{\text{irs}_\text{D}} $ is the combined $z$th spatial frequency. \al{Defining 
\begin{align}
\bof{m}_y = \bof{D}(\bof{\alpha})\bof{\Omega}_y^\R{T}\bof{n}_y(\mu_{\text{y}}),\quad \bof{m}_z = \bof{D}(\bof{\beta})\bof{\Omega}_z^\R{T}\bof{n}_z(\psi_{\text{z}})  \label{eq:mymz}
\end{align}
we can rewrite} \eqref{X_y} and \eqref{X_z} as equivalent rank-one matrices in compact form respectively, as
\begin{align}
&\al{\hat{\bof{U}}_y} = \left(\bof{a}_y(\mu_{\text{bs}}) \otimes \bof{q}_y(\mu_{\text{ue}})\right)\bof{m}_y^\R{T},\label{final_X_y_tilde}\\
&\al{\hat{\bof{U}}_z} = \left(\bof{a}_z(\psi_{\text{bs}}) \otimes  \bof{q}_z(\psi_{\text{ue}})\right)\bof{m}^\R{T}_z. \label{final_X_z_tilde}
\end{align} 

\subsubsection{\al{Estimation of BS and UE channel parameters}}
Equations \eqref{final_X_y_tilde} and \eqref{final_X_z_tilde} can be recast in equivalent tensor forms. Let us define the third-order tensors
\al{$\mathcal{\hat{U}}_y \in \mathbb{C}^{Q_y \times M_y \times K}$ and $\mathcal{\hat{U}}_z \in \mathbb{C}^{Q_z \times M_z \times K}$, respectively, \textit{via} the mappings  $[\mathcal{\hat{U}}_y]_{q_y,m_y,k} \doteq [\hat{\bof{U}}_y]_{(q_y-1)M_y+m_y,k}$ and $[\mathcal{\hat{U}}_z]_{q_z,m_z,k} \doteq [\hat{\bof{U}}_z]_{(q_z-1)M_z+m_z,k}$,
where $q_t=1, \ldots, Q_t$, $m_t=1,\ldots, M_t$, $t \in \{y,z\}$, and $k=1, \ldots, K$. 
Due to their rank-one structure, these tensors can be expressed as
$\mathcal{\hat{U}}_y =\bof{q}_y(\mu_{\text{ue}})\circ\ \bof{a}_y(\mu_{\text{bs}})\circ\bof{m}_y$ and $\mathcal{\hat{U}}_z =\bof{q}_z(\mu_{\text{ue}})\circ\ \bof{a}_yz(\mu_{\text{bs}})\circ\bof{m}_z$. Hence, the $y$-domain and $z$-domain channel parameters can be estimated by solving the following rank-one tensor approximation problems }
\begin{equation}
\min _{\bof{a}_y,\bof{q}_y,\bof{m}_y} \left\| \al{\mathcal{\hat{U}}_y} - \bof{q}_y(\mu_{\text{ue}})\circ\ \bof{a}_y(\mu_{\text{bs}})\circ\bof{m}_y \right\|_\R{F}^2, \label{tensorcost_y}
\end{equation}
\begin{equation}
\min _{\bof{a}_z,\bof{q}_z,\bof{m}_z} \left\| \al{\mathcal{\hat{U}}_z} - \bof{q}_z(\psi_{\text{ue}})\circ\ \bof{a}_z(\psi_{\text{bs}})\circ\bof{m}_z \right\|_\R{F}^2. \label{tensorcost_z}
\end{equation}
To solve \eqref{tensorcost_y} and \eqref{tensorcost_z}, the truncated higher order singular value decomposition (HOSVD) \cite{Kolda_2009} or the tensor-power-method detector (TPMD) \cite{Asim_2020} algorithms can be applied to obtain the estimates $\hat{\bof{a}}_y(\mu_{\text{bs}})$, $\hat{\bof{q}}_y(\mu_{\text{ue}})$,  $\hat{\bof{a}}_z(\psi_{\text{bs}})$, and
$\hat{\bof{q}}_z(\psi_{\text{ue}})$. More specifically, for both algorithms, the $y$th and $z$th domain components are found as the dominant $n$-mode  singular vectors of \al{$\mathcal{\mathcal{\hat{U}}}_y$  and $\mathcal{\mathcal{\hat{U}}}_z$,} respectively. For more details on the HOSVD and TPMD algorithms, we refer the reader to \cite{Kolda_2009, Asim_2020}. To extract the associated spatial frequencies, $\mu_{\text{bs}}$, $\mu_{\text{ue}}$, $\psi_{\text{bs}}$, and $\psi_{\text{ue}}$,  a grid peak-search approach is used, i.e.
\begin{align}
\hat{\mu}_{\text{bs}} &= \arg\max_{\mu_{\text{bs}}}\left\{\hat{\bof{a}}_y^\R{H}(\mu_{\text{bs}})\bof{a}_y(\mu_{\text{bs}})\right\} \label{mu_est_grid}\\
\hat{\mu}_{\text{ue}} &= \arg\max_{\mu_{\text{ue}}}\left\{\hat{\bof{q}}_y^\R{H}(\mu_{\text{ue}})\bof{a}_y(\mu_{\text{ue}})\right\} \\
\hat{\psi}_{\text{bs}} &= \arg\max_{\psi_{\text{bs}}}\left\{\hat{\bof{a}}_z^\R{H}(\psi_{\text{bs}})\bof{a}_z(\psi_{\text{bs}})\right\} \\ \hat{\psi}_{\text{ue}}&= \arg\max_{\psi_{\text{ue}}}\left\{\hat{\bof{q}}_z^\R{H}(\psi_{\text{ue}})\bof{a}_z(\psi_{\text{ue}})\right\}  \label{psi_est_grid}
\end{align}

\subsubsection{\al{Estimation of the IRS-related channel parameters}}
\al{Once the BS and the UE channel parameters are found, the IRS-related channel steering vectors can be estimated from a ``filtered'' version of the horizontal and vertical cascaded channel tensors using transmit and receive beamforming. Defining these beamforming vectors as
\begin{equation}
\bof{f}_y = \bof{a}_y^\ast(\hat{\mu}_{\text{bs}}), \,
\bof{f}_z = \bof{a}_z^\ast(\hat{\psi}_{\text{bs}}), \,
\bof{g}_y = \bof{q}_y^\ast(\hat{\mu}_{\text{ue}}), \,
\bof{g}_z = \bof{q}_z^\ast(\hat{\psi}_{\text{ue}}),\nonumber
\end{equation} 
we get
\begin{align}
\bof{l}_y&= \mathcal{\hat{U}}_y \times_1 \bof{g}_y \times_2 \bof{f}_y = \gamma_y\bof{m}_y\in \mathbb{C}^{K \times 1},\label{eq:filteringly}\\
\bof{l}_z&= \mathcal{\hat{U}}_z \times_1 \bof{g}_z \times_2 \bof{f}_z = \gamma_z\bof{m}_z,
\in \mathbb{C}^{K \times 1}\label{eq:filteringlz}
\end{align}
where $\gamma_y$ and $\gamma_z$ are scalars representing the beamforming gains along $y$ and $z$ domains, respectively.}
In the following, we detail the steps associated with the estimation of the IRS steering vectors $\bof{n}_y(\mu_y)$ and $\bof{n}_z(\psi_z)$. To this end, \al{let us define $\bof{l}= \bof{l}_y \odot \bof{l}_z= (\bof{l}^\R{T}_y \diamond \bof{l}^\R{T}_z)^\R{T}$. Using the expressions in (\ref{eq:mymz}) and considering} the properties $\bof{AB} \diamond \bof{CD} = \left( \bof{A} \otimes \bof{C}\right) \left(\bof{B} \diamond \bof{D}\right)$, and $\left(\bof{A}\bof{D}_\text{A}\right) \diamond \left(\bof{B}\bof{D}_\text{B}\right) = \left( \bof{A} \diamond \bof{B}\right) \left(\bof{D}_\text{A} \odot \bof{D}_\text{B}\right)$ yields
\begin{align}
\bof{l}^\R{T} 
&= \big[\gamma_y\bof{n}^\R{T}_y(\mu_{\text{y}})\bof{\Omega}_y \bof{D}(\bof{\alpha})\big] \diamond \big[\gamma_z\bof{n}^\R{T}_z(\psi_{\text{z}})\bof{\Omega}_z \bof{D}(\bof{\beta})\big] \nonumber \\
&= 	\gamma_y\gamma_z \big[\bof{n}^\R{T}_y(\mu_{\text{y}}) \otimes \bof{n}^\R{T}_z(\psi_{\text{z}}) \big] \big[ \left( \bof{\Omega}_y\bof{D}(\bof{\alpha})\right) \diamond  \left(\bof{\Omega}_z\bof{D}(\bof{\beta}) \right) \big] \nonumber\\
&= \gamma_y\gamma_z\big(\bof{n}^\R{T}_y(\mu_{\text{y}}) \otimes \bof{n}^\R{T}_z(\psi_{\text{z}}) \big) \big(\bof{\Omega}_y \diamond \bof{\Omega}_z\big) \big( \bof{D}(\bof{\alpha}) \odot \bof{D}(\bof{\beta})\big), \nonumber 
\end{align}
\al{Since the scaling matrices cancel each other (see (\ref{X_tilde_est_y})-(\ref{X_tilde_est_z})), we have $\bof{D}(\bof{\alpha}) \odot \bof{D}(\bof{\beta})=\bof{I}_K$, yielding}
\begin{align}
\bof{l} = \bof{\Omega}^\R{T}\underbrace{\gamma\left(\bof{n}_y(\mu_{\text{y}}) \otimes \bof{n}_z(\psi_{\text{z}})\right)}_{\bof{n}(\varphi)} \label{second_last_v(varphi)}
\end{align} 
\fa{where $\gamma = \gamma_y\gamma_z$. Assuming $\bof{\Omega}^* \bof{\Omega}^\R{T} =\bof{I}_{N}$, an estimate of $\bof{n}(\varphi) \in \mathbb{C}^{N \times 1}$ can be obtained by means of matched filtering with the IRS training matrix, i.e.,:}
\begin{align}
\hat{\bof{n}}(\varphi) = \al{\bof{\Omega}^\ast\bof{l}. 
\label{n_hat}}
\end{align} 
\al{Due to the separable structure of  $\hat{\bof{n}}(\varphi)$ in (\ref{second_last_v(varphi)})}, decoupled estimates $\hat{\bof{n}}_y(\mu_{\text{y}})$ and $\hat{\bof{n}}_z(\psi_{\text{z}})$ can be found from the following rank-one matrix approximation problem 
\begin{align}
\left\{\hat{\bof{n}}_y(\mu_\text{y}),\hat{\bof{n}}_z(\psi_{\text{z}})\right\}= \arg\min_{\bof{}} \left \|\hat{\bof{N}} - \bof{n}_z(\psi_{\text{z}}) \bof{n}_y^\R{T}(\mu_{\text{y}})  \right\|_\R{F}^2,\nonumber
\end{align}
\al{where $\hat{\bof{N}} = \R{unvec}_{(N_z \times N_y)}\left\{\hat{\bof{n}}(\varphi) \right\} \in \mathbb{C}^{N_z \times N_y}.$}
These estimates are then found as $\hat{\bof{n}}_z(\psi_{\text{z}}) = \sqrt{\sigma_1}\bof{u}_1$ and $\hat{\bof{n}}_y(\mu_{\text{y}}) = \sqrt{\sigma_1}\bof{v}_1^\ast$,
where $\bof{u}_1$ and $\bof{v}_1$ are the dominant left and right singular vectors of $\hat{\bof{N}}$.
Adopting the same procedure as \eqref{mu_est_grid}-\eqref{psi_est_grid}, the spatial frequencies $\mu_{\text{y}}$ and $\psi_{\text{z}}$ are extracted as 
\begin{align}
\hat{\mu}_{\text{y}} &= \arg\max_{\mu_{\text{y}}}\left\{\hat{\bof{n}}_y^\R{H}(\mu_{\text{y}})\bof{a}_y(\mu_{\text{y}})\right\} \label{mu_y_est}\\
\hat{\psi}_{\text{z}} &= \arg\max_{\psi_{\text{z}}}\left\{\hat{\bof{n}}_z^\R{H}(\psi_{\text{z}})\bof{a}_z(\psi_{\text{z}})\right\}. \label{psi_z_est}
\end{align}
Note that \eqref{mu_est_grid}-\eqref{psi_est_grid} and \eqref{mu_y_est}-\eqref{psi_z_est} provide the estimates of all the involved quantities. A detailed explanation of the HKMR method is given in Algorithm \ref{alg:HKMR}.
\begin{algorithm}[tb]
	{\small
		\caption{\label{alg:HKMR} Hybrid Kronecker factorization and Multi-Rank-one approximations (HKMR) Method}
				\Begin{
			\For{$k = 1\; \textrm{to} \; K$}{Estimate $\hat{\bof{X}}_y^{(k)}$ and $\hat{\bof{X}}_z^{(k)}$ from $\bof{X}_{k}$ \textit{via} the LS\\ 
					\nonl Kronecker factorization problem in \eqref{LS_kron_probel} \label{LSkron_alg_1}\\ Obtain $\hat{\bof{U}}_y^{(k)}\leftarrow \hat{\bof{X}}_y^{(k)}\bof{S}_y^\R{H}$ and $\hat{\bof{U}}_z^{(k)} \leftarrow \hat{\bof{X}}_z^{(k)}\bof{S}_z^\R{H}$ \\ 
					\nonl by temporal filtering with the known pilots}
		
			$\hat{\bof{U}}_y = \big[\R{vec}\{\hat{\bof{U}}_y^{(1)}\},\dots,\R{vec}\{\hat{\bof{U}}_y^{(K)}\} \big]$ \\
		$\hat{\bof{U}}_z = \big[\R{vec}\{\hat{\bof{U}}_z^{(1)}\},\dots,\R{vec}\{\hat{\bof{U}}_z^{(K)}\} \big]$ \\
		Build $\mathcal{\hat{U}}_y$ and $\mathcal{\hat{U}}_z$ by reshaping $\hat{\bof{U}}_y$ and $\hat{\bof{U}}_z$ \\
		Find the BS and UE channel steering vectors: \\
		\nonl
		$\{\hat{\bof{a}}_y,\hat{\bof{q}}_y\} \!\leftarrow \!\R{HOSVD}\big(\mathcal{\hat{U}}_y\big)$,
		$\{\hat{\bof{a}}_z,\hat{\bof{q}}_z\} \! \leftarrow \!\R{HOSVD}\big(\mathcal{\hat{U}}_z\big)$ \label{alg_1_UyUz} \\
		Using the BS and UE beamforming vectors \nonl \\
		\nonl $\bof{f}_y\!=\! \bof{a}_y^\ast(\hat{\mu}_{\text{bs}})$, $\bof{f}_z\!=\!\bof{a}_z^\ast(\hat{\psi}_{\text{bs}})$, $\bof{g}_y\!=\!\bof{q}_y^\ast(\hat{\mu}_{\text{ue}})$, $\bof{g}_z\!=\!\bof{q}_z^\ast(\hat{\psi}_{\text{ue}})$ \\
		\nonl obtain $\{\bof{l}_y,\bof{l}_z\}$ from 
		$\{\mathcal{\hat{U}}_y,\mathcal{\hat{U}}_z\}$ by spatial filtering  \label{alg_1_bf_comb} 
		\\
		Build $\bof{l}= \bof{l}_y \odot \bof{l}_z$ and  obtain $\hat{\bof{n}}(\varphi)=\bof{\Omega}^*\bof{l}$.
		\\
		From $\hat{\bof{N}} = \R{unvec}_{(N_z \times N_y)}\left\{\hat{\bof{n}}(\varphi) \right\}$
		find the IRS \\  \nonl  steering vectors:
			$\left\{\hat{\bof{n}}_y,\hat{\bof{n}}_z\right\} \leftarrow \R{SVD}\big(\hat{\bof{N}}\big)$ \label{alg_1_N} 
			\\
			Find the channel angular parameters according to \\ \nonl equations \eqref{mu_est_grid}-\eqref{psi_est_grid} and \eqref{mu_y_est}-\eqref{psi_z_est}
	}}
\end{algorithm}

\subsection{Two-stage Higher-dimensional Rank-one Approximations (TSHDR) Method}
\al{As shown in the previous section, the HKMR method starts by decoupling the received pilots into their horizontal and vertical components using the $y$ and $z$ domain pilot signals. Here, two-dimensional channel decoupling is done at a later stage, which allows estimating the BS, UE, and IRS channel parameters jointly by means of two rank-one tensor approximation problems.}
Recalling \eqref{recived_eq} and applying right-filtering with the known pilot matrix yields \al{an estimate of the cascaded MIMO channel at the $k$-th block, as}  
\begin{align}
\al{\bof{U}_k} = \bof{X}_k\bof{S}^\R{H} =  \bof{G} \R{diag}\left\{\bof{\omega}_{k}\right\}\bof{H} + \al{\bof{V}'_k} \in \mathbb{C}^{Q \times M}, \label{MF_tshdr}
\end{align}
where we have $\bof{S}\bof{S}^\R{H} = \bof{I}_M$ \al{and $\bof{V}'_k$ is the filtered noise term.}  Applying the $\R{vec}\left\{.\right\}$ operator leads to
\begin{align}
\bof{u}_k = \R{vec}\left\{\bof{U}_k\right\} = \big(\bof{H}^\R{T} \diamond \bof{G}\big)\bof{\omega}^{(k)} \al{+ \bof{v}'_k}\in \mathbb{C}^{QM \times 1}. \nonumber
\end{align}
Collecting the $K$ received vectors results in
\begin{align}
\bof{U} = \big[\bof{u}_1,\dots, \bof{u}_K\big] =  \big(\bof{H}^\R{T} \diamond \bof{G}\big)\bof{\Omega} \al{+ \bof{V}'} \in \mathbb{C}^{QM \times K}, \nonumber 
\end{align}
where $\bof{V}' = \big[\bof{v}'_1,\dots, \bof{v}'_K\big] \in \mathbb{C}^{QM \times K}$. By applying MF with the known IRS matrix, we get
\begin{align}
\bof{E} \al{ =\bof{U}\bof{\Omega}^\R{H}}= \bof{H}^\R{T} \diamond \bof{G} \al{+ \bof{V}''}\in\mathbb{C}^{QM \times N}.
\label{Khatri-Rao channel}
\end{align}
\al{As shown in \cite{Gilderlan_2021}, the individual channel matrices $\bof{H}$ and $\bof{G}$ can be estimated from the pilot-filtered signal matrix $\hat{\bof{E}}$ by solving the following} LS Khatri-Rao factorization problem 
\begin{align}
\hat{\bof{H}},\hat{\bof{G}} = \arg\min\left\| \hat{\bof{E}} - \bof{H}^\R{T} \diamond \bof{G} \right\|_\R{F}^2. \label{ls_opt}
\end{align}
\al{Taking into account} the Kronecker factorization structure of the channels $\bof{H}$ and $\bof{G}$ defined in \eqref{final_H}-\eqref{final_G}, we can expand problem \eqref{ls_opt} to the following one
\begin{align}
\hspace{-2ex}\hat{\bof{H}},\hat{\bof{G}} = \arg\min\left\| \hat{\bof{E}} - \left(\bof{H}_y \otimes \bof{H}_z\right)^\R{T} \diamond \left(\bof{G}_y \otimes \bof{G}_z\right) \right\|_\R{F}^2. \label{LS-Khatri}
\end{align}
\al{Let us consider the following property}
\begin{equation}
\left(\bof{A} \otimes \bof{B}\right) \diamond \left(\bof{C} \otimes \bof{D}\right) = \bof{P} \left[\left(\bof{A} \diamond \bof{C}\right) \otimes \left(\bof{B} \diamond \bof{D}\right)\right]\nonumber
\end{equation} 
\al{where $\bof{A} \in \mathbb{C}^{I\times R}, \bof{B} \in \mathbb{C}^{J\times S}, \bof{C} \in \mathbb{C}^{K\times R}, \bof{D} \in \mathbb{C}^{L\times S}$, and $\bof{P} \in \mathbb{R}^{LJKI\times LKJI}$ is a block permutation matrix. Applying this property to (\ref{LS-Khatri}) with the correspondences $(\bof{A},\bof{B},\bof{C},\bof{D})\longleftrightarrow (\bof{H}^\R{T}_y,\bof{H}^\R{T}_z,\bof{G}_y,\bof{G}_z)$, and $(I,J,K,L,R,S)\longleftrightarrow(M_y,M_z,Q_y,Q_z,N_y,N_z)$
(see \cite{fazaleasim2023tensorbased} for details on the structure of this matrix), and } defining $\bof{J} =\bof{P}\bof{E}$, we can recast (\ref{LS-Khatri})  the following problem 
\begin{align}
\hspace{-2ex}\hat{\bof{H}},\hat{\bof{G}} = \arg\min\left\| \bof{J} - \left(\bof{H}_y^\R{T} \diamond \bof{G}_y\right) \otimes \left(\bof{H}_z^\R{T} \diamond \bof{G}_z\right) \right\|_\R{F}^2. \label{new_cost_function}
\end{align}

\al{Let us define $\bof{J}_y = \bof{H}_y^\R{T} \diamond \bof{G}_y \in \mathbb{C}^{Q_yM_y \times N_y}$ and  $\bof{J}_z = \bof{H}_z^\R{T} \diamond \bof{G}_z \in \mathbb{C}^{Q_zM_z \times N_z}$ as the combined Khatri-Rao channels associated with the horizontal and vertical domains, respectively. Capitalizing again on the LS Kronecker factorization, and analogously to \eqref{LS_kron_probel}, we can estimate $\bof{J}_y$ and $\bof{J}_z$ (up to scaling) by solving the problem
\begin{equation}
   \{\bof{J}_y,\bof{J}_z\}= \arg\min_{\bof{}} \left \|\bof{J} - \bof{J}_y \otimes \bof{J}_z  \right\|_\R{F}^2,\label{eq:lskronJ}
\end{equation}
the solution of which leads to $\hat{\bof{J}}_y= \alpha \bof{J}_y $ and $\hat{\bof{J}}_z= \beta\bof{J}_z$ where $\alpha$ and $\beta$ are scaling factors with $\alpha\beta=1$.}
\al{Defining $\bof{\hat{j}}_y =\R{vec}\big\{\hat{\bof{J}}_y\big\}=\R{vec}\big\{\big[\bof{H}_y^\R{T} \diamond \bof{G}_y\big]\alpha \big\} \in \mathbb{C}^{Q_yM_yN_y \times 1}$ and $\bof{\hat{j}}_z =\R{vec}\big\{\hat{\bof{J}}_z\big\}=\R{vec}\big\{\big[\bof{H}_z^\R{T} \diamond \bof{G}_z \big] \beta\big\}  \in \mathbb{C}^{Q_zM_zN_z \times 1}$, and taking into account the structures of $\{\bof{H}_y,\bof{H}_z\}$ and  $\{\bof{G}_y,\bof{G}_z\}$ given in (\ref{final_H}) and (\ref{final_G}), respectively, we obtain}
\begin{align}
\bof{\hat{j}}_y=\overline{\bof{n}}_y(\mu_{\text{y}}) \otimes \bof{a}_y(\mu_{\text{bs}}) \otimes  \bof{q}_y(\mu_{\text{ue}})  \label{j_y_hat}  \\
\bof{\hat{j}}_z =\overline{\bof{n}}_z(\psi_{\text{z}}) \otimes \bof{a}_z(\psi_{\text{bs}}) \otimes  \bof{q}_z(\psi_{\text{ue}}) \label{j_z_hat}
\end{align}
where $\overline{\bof{n}}_y(\mu_{\text{y}}) = \alpha \bof{n}_y(\mu_{\text{y}})$, and $\overline{\bof{n}}_z(\psi_{\text{z}}) = \beta\bof{n}_z(\psi_{\text{z}})$. 
Let us define the third-order tensors
\al{$\mathcal{\hat{J}}_y \in \mathbb{C}^{Q_y \times M_y \times N_y}$ and $\mathcal{\hat{J}}_z \in \mathbb{C}^{Q_z \times M_z \times N_z}$, respectively, \textit{via} the mappings  $[\mathcal{\hat{J}}_y]_{q_y,m_y,n_y} \doteq [\hat{\bof{J}}_y]_{(m_y-1)Q_y+q_y,n_y}$ and $[\mathcal{\hat{J}}_z]_{q_z,m_z,n_z} \doteq [\hat{\bof{J}}_z]_{(m_z-1)Q_z+q_z,n_z}$,
where $q_t=1, \ldots, Q_t$, $m_t=1,\ldots, M_t$, $n_t=1,\ldots, N_t$, $t \in \{y,z\}$.
Due to their rank-one structure, we can find the channel parameters directly from $\mathcal{\hat{J}}_y$ and $\mathcal{\hat{J}}_z$ by solving horizontal and vertical rank-one tensor approximation problems as follows}
\begin{equation}
\hat{\bof{a}}_y,\hat{\bof{q}}_y,\hat{\overline{\bof{n}}}_y=\min _{\bof{a}_y,\bof{q}_y,\bof{n}_y^\prime} \left\|  \hat{\mathcal{J}}_y - \bof{q}_y(\mu_{\text{ue}})\circ\ \bof{a}_y(\mu_{\text{bs}})\circ\overline{\bof{n}}_y(\mu_{\text{y}}) \right\|_\R{F}^2, \label{tshdr_y}
\end{equation}
\begin{equation}
\hat{\bof{a}}_z,\hat{\bof{q}}_z,\hat{\overline{\bof{n}}}_z=\min _{\bof{a}_z,\bof{q}_z,\bof{n}_z^\prime} \left\|  \hat{\mathcal{J}}_z  - \bof{q}_z(\psi_{\text{ue}})\circ\ \bof{a}_z(\psi_{\text{bs}})\circ \overline{\bof{n}}_z(\psi_{\text{z}}) \right\|^2_\R{F}. \label{tshdr_z}
\end{equation}
\al{Note that} problems \eqref{tshdr_y}-\eqref{tshdr_z} can be solved similarly to problems \eqref{tensorcost_y}-\eqref{tensorcost_z} using, for instance, the truncated HOSVD algorithm,  \al{as discussed in the previous section.} The difference here is that the effective IRS steering vectors are estimated jointly with the \al{BS and UE} steering vectors.
After solving \eqref{tshdr_y} and \eqref{tshdr_z}, the spatial frequencies can be extracted according to \eqref{mu_est_grid}-\eqref{psi_est_grid} and \eqref{mu_y_est}-\eqref{psi_z_est}. \al{The summary of the TSHDR algorithm is provided in Algorithm \ref{alg:TSHDR}.}
\begin{algorithm}[tb]
{\small
	\caption{\label{alg:TSHDR} Two-stage Higher-Dimensional Rank-one approximations (TSHDR) Method}
	\Begin{
	\For{$k=1\; \textrm{to} \; K$}{\al{Estimate the cascaded MIMO channel \\ \nonl in the LS sense: $\bof{U}_k = \bof{X}_k\bof{S}^\R{H}$ \\ 
  }}
  \al{$\bof{U}= \big[\R{vec}\{\bof{U}_1\}, \ldots, \R{vec}\{\bof{U}_K\}\big]$} \\ 
	\al{Obtain $\bof{E} = \bof{U}\bof{\Omega}^\R{H}$ and build $\bof{J}=\bof{P}\bof{E}$ by permutation} \\	
 \al{Find the estimates $\hat{\bof{J}}_y$ and $\hat{\bof{J}}_z$ from $\bof{J}$ \textit{via} the LS\\ \nonl Kronecker factorization problem in \eqref{eq:lskronJ}} \label{LS_algo_2}\\
	 \al{Build $\mathcal{\hat{J}}_y$ and $\mathcal{\hat{J}}_z$ by reshaping $\hat{\bof{J}}_y$ and $\hat{\bof{J}}_z$} \\	
  \al{Find the BS, UE, and IRS channel steering vectors}\\
  \nonl \al{$\big\{\hat{\bof{a}}_y,\hat{\bof{q}}_y,\hat{\overline{\bof{n}}}_y\big\}\leftarrow \R{HOSVD}\big( \hat{\mathcal{J}}_y\big)$} \\
 \nonl \al{$\big\{\hat{\bof{a}}_z,\hat{\bof{q}}_z,\hat{\overline{\bof{n}}}_z\big\}\leftarrow \R{HOSVD}\big(\hat{\mathcal{J}}_z \big)$} \label{algJ_yJ_z}\\
  \al{Find the channel angular parameters according to \\ \nonl equations \eqref{mu_est_grid}-\eqref{psi_est_grid} and \eqref{mu_y_est}-\eqref{psi_z_est}} 
 }
 }
\end{algorithm}
\subsection{Computational Complexity Analysis}
The complexity of the HKMR method is calculated in multiple steps. The first step has a complexity equivalent to that of a rank-one matrix approximation in \eqref{LS_Kron_prob_A}. Since this step is repeated $K$ times, the total complexity is $\mathcal{O}\left(KQ^2T\right)$. In the second step, MF is applied to both $y$th and $z$th domains i.e., \eqref{MF_y} and \eqref{MF_z}, respectively, whose overall complexity is $K$ times the complexity of the individual $y$-domain and $z$-domain matrix products, which results in $\mathcal{O}\left(K\left(Q_yT_yM_y+Q_zT_zM_z\right)\right)$. The total complexity of the HKMR method is then given by
\begin{align}
\Gamma_{\text{HKMR}} = \mathcal{O}\,K\left(Q^2T+Q_yT_yM_y+Q_zT_zM_z+Q_y^2M_y  
 +M_y^2Q_y+Q_z^2M_z+M_z^2Q_z\right)+N_z^2N_y. \label{hkmr_complexty}
\end{align}
According to \eqref{hkmr_complexty}, the complexity of the HKMR method increases proportionally to the squared size of the number of the transmit antennas, the number of receive antennas, and the number of reflecting elements at the IRS.
\\ The complexity of TSHDR method can also be calculated using multiple steps. The first step consists of $K$ matrix products associated with the MF operations applied to \eqref{MF_tshdr}, yielding a complexity of $\mathcal{O}\left(KMQT\right)$. The total complexity of the TSHDR method, including the multiple rank-one approximations in \eqref{tshdr_y}, \eqref{tshdr_z}, is given by
\begin{align}
&\Gamma_{\text{TSHDR}} = \mathcal{O} \left(KMQT + M^2Q^2N + Q_y^2M_yN_y+M_y^2Q_yN_y \right.\nonumber \\
&\left. +N_y^2Q_yM_y+Q_z^2M_zN_z+M_z^2Q_zN_z+N_z^2Q_zM_z\right). \label{tshdr_complexty}
\end{align}
Based on \eqref{tshdr_complexty}, the complexity of the TSHDR method is proportional to the squared size of the number of the transmit antennas, the number of receive antennas, and the number of reflecting elements at the IRS.
\\
The complexity of the HDR method is $\mathcal{O}\left(Q^2MNTK + QMN\left(Q_z+Q_y+M_z+M_y+N_z+N_y\right)\right)$ \cite{fazaleasim2023tensorbased}.
 Note that the complexity of the KRF method \cite{Gilderlan_2021} is given by $\mathcal{O}\left(Q^2MNTK+N^2Q^2M^2\right)$. Also, the complexity of the LS method \cite{Gilderlan_2021} is given as $\mathcal{O}\left(Q^2MNTK\right)$.

The benefits of the HKMR and TSHDR methods, where the parameter estimation problem is decoupled into horizontal and vertical sub-problems, is the reduced overall computational complexity from multiplication ($N_yN_z$) to addition ($N_y+N_z$) \al{in their processing steps}. Moreover, these methods enable the parallelization of the horizontal and vertical estimations, which is not possible with classical state-of-the-art approaches.

\section{Cramér-Rao Lower Bound}
The CRLB is derived in this section \al{as a reference for comparisons} of the proposed algorithms. The CRLB for an unbiased estimator is found from the inverse of the diagonal of the Fisher information matrix (FIM) \cite{Asim_2021}. 
Two FIM matrices are found separately for both \gf{the $y$ and the $z$ domains} 
\begin{align}
\R{var}(\hat{\eta}_y^{(i)}) \ge \left[\bof{F}^{-1}(\bof{\eta}_y)\right]_{ii},\quad
\R{var}(\hat{\eta}_z^{(i)}) \ge \left[\bof{F}^{-1}(\bof{\eta}_z)\right]_{ii},\nonumber
\end{align}
where $\bof{\eta}_y = \left[\mu_{\text{bs}},\mu_{\text{ue}},\mu_{\text{y}}\right]$ and $\bof{\eta}_z = \left[\psi_{\text{bs}},\psi_{\text{ue}},\psi_{\text{z}}\right]$.The lower bound on the standard deviation of the estimation error is 
\begin{align}
\sqrt{\R{CRLB}(\hat{\eta}_y^{(i)})} = \sqrt{\left[\bof{F}^{-1}(\bof{\eta}_y)\right]_{ii}},\nonumber
\end{align} 
where $ii$ is the diagonal element of $\bof{F}(\bof{\eta}_y)$ and $\bof{F}(\bof{\eta}_z)$. Then, $\bof{F}(\bof{\eta}_y)$ and $\bof{F}(\bof{\eta}_z)$ are given as 
\begin{align}
\left[\B{F}(\bof{\eta}_y)\right]_{ij} =
\frac{2}{\sigma_n^2}\,\, \R{Re}\,\,\left\{\R{tr}\left\{\frac{\partial \bof{S}^\R{H}(\bof{\eta}_y)}{\partial \eta_{y}^{(i)}}\,\,\frac{\partial\bof{S}(\bof{\eta}_y)}{\partial\eta_{y}^{(j)}}\right\}\right\}, \nonumber
\end{align}
\begin{align}
\left[\B{F}(\bof{\eta}_z)\right]_{ij} =
\frac{2}{\sigma_n^2}\,\, \R{Re}\,\,\left\{\R{tr}\left\{\frac{\partial \bof{S}^\R{H}(\bof{\eta}_z)}{\partial \eta_{z}^{(i)}}\,\,\frac{\partial\bof{S}(\bof{\eta}_z)}{\partial\eta_{z}^{(j)}}\right\}\right\}, \nonumber
\end{align}
where $\bof{S}(\bof{\eta}_y)$ and $\bof{S}(\bof{\eta}_z)$ have the following structure
\begin{align}
\bof{S}(\bof{\eta}_y) = \bof{H}_y^\R{T} \diamond \bof{G}_y,\quad 
\bof{S}(\bof{\eta}_z) = \bof{H}_z^\R{T} \diamond \bof{G}_z.\nonumber
\end{align}
These matrices can be \al{respectively} rewritten as 
\begin{align*}
\bof{S}(\bof{\eta}_y) = \left(\bof{a}_y(\mu_{\text{bs}}) \otimes  \bof{q}_y(\mu_{\text{ue}})\right) \bof{n}_y^\R{T}(\mu_{\text{y}}), \\
\bof{S}(\bof{\eta}_z) = \left(\bof{a}_z(\psi_{\text{bs}}) \otimes  \bof{q}_z(\psi_{\text{ue}})\right) \bof{n}_z^\R{T}(\psi_{\text{z}}).\nonumber
\end{align*}
\begin{figure*}
	\centering
	\begin{subfigure}[b]{0.45\textwidth}
		\centering
		\includegraphics[width=\textwidth]{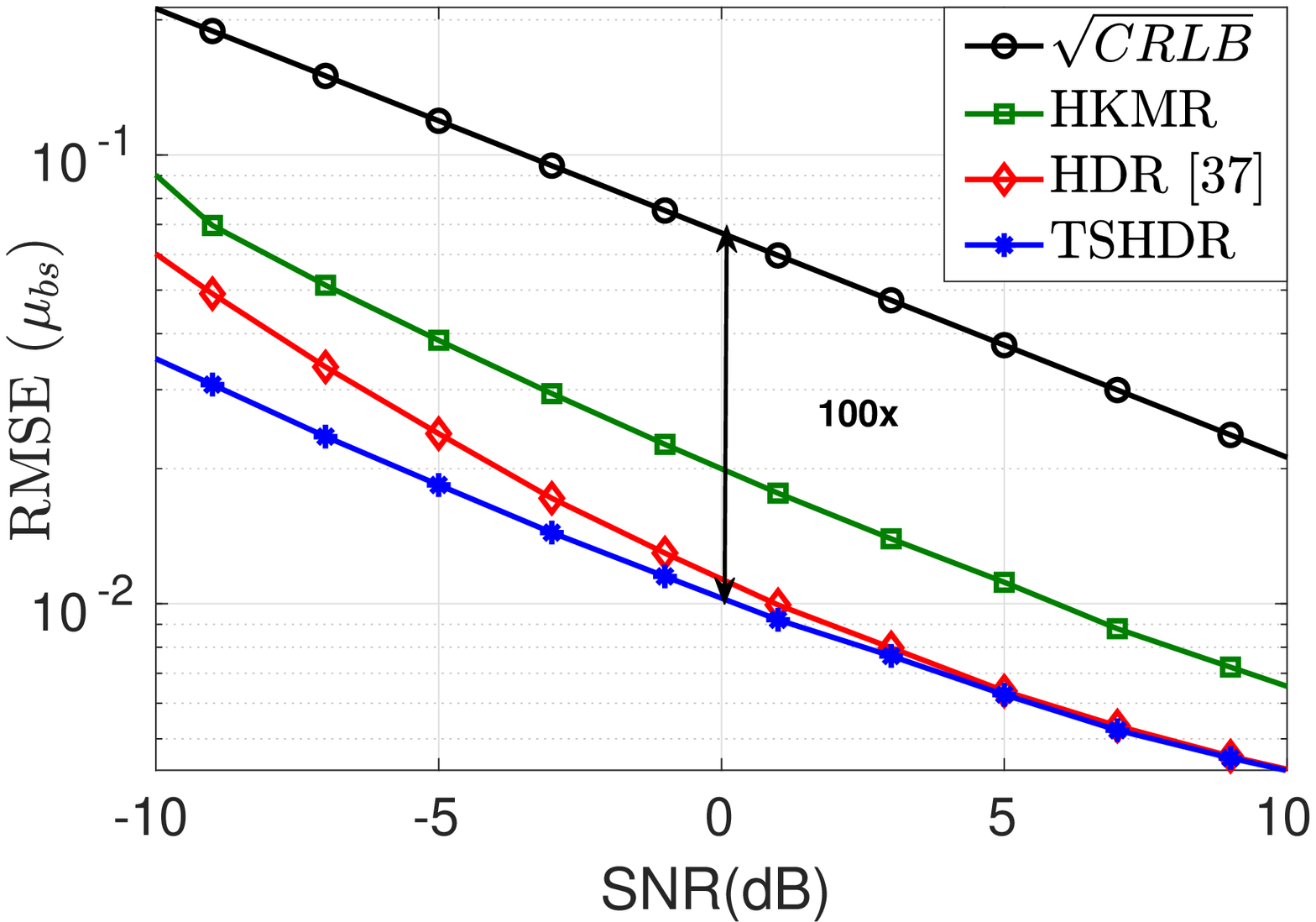}
		\caption{$\R{RMSE}$ of the $y$th spatial frequency $\mu_{\text{bs}}$}
		\label{fig:3_mu_bs}
	\end{subfigure}
	\begin{subfigure}[b]{0.45\textwidth}
		\centering
		\includegraphics[width=\textwidth]{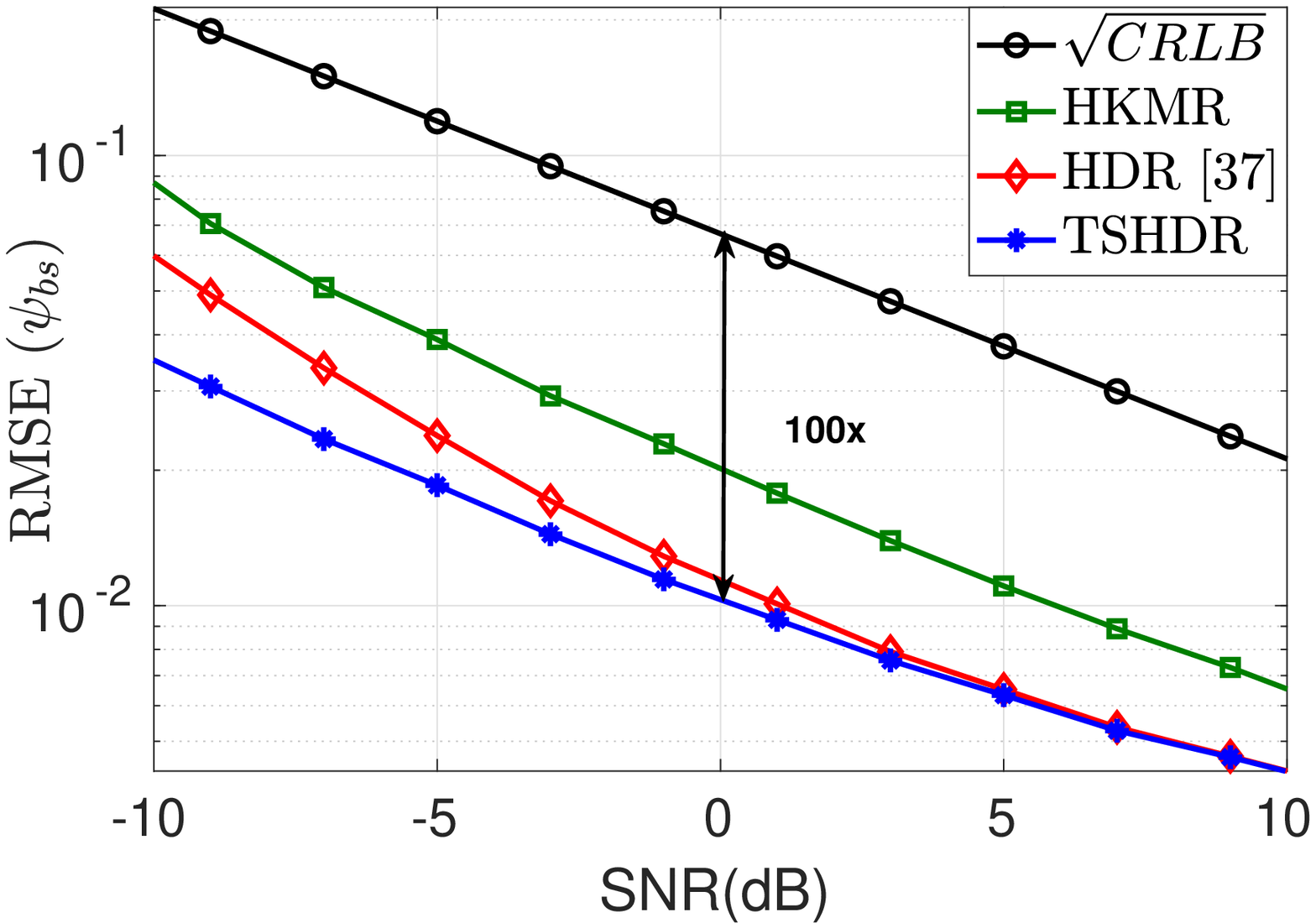}
		\caption{$\R{RMSE}$ of the $z$th spatial frequency $\psi_{\text{bs}}$}
		\label{fig:4_psi_bs}
	\end{subfigure}
	\caption{$\R{RMSE}$ for the estimated transmit spatial frequencies associated with the BS assuming $M=16$, $Q=16$, and $N=16$.}
	\label{fig:bs}
\end{figure*}
\begin{figure*}
	\centering
	\begin{subfigure}[b]{0.45\textwidth}
		\centering
		\includegraphics[width=\textwidth]{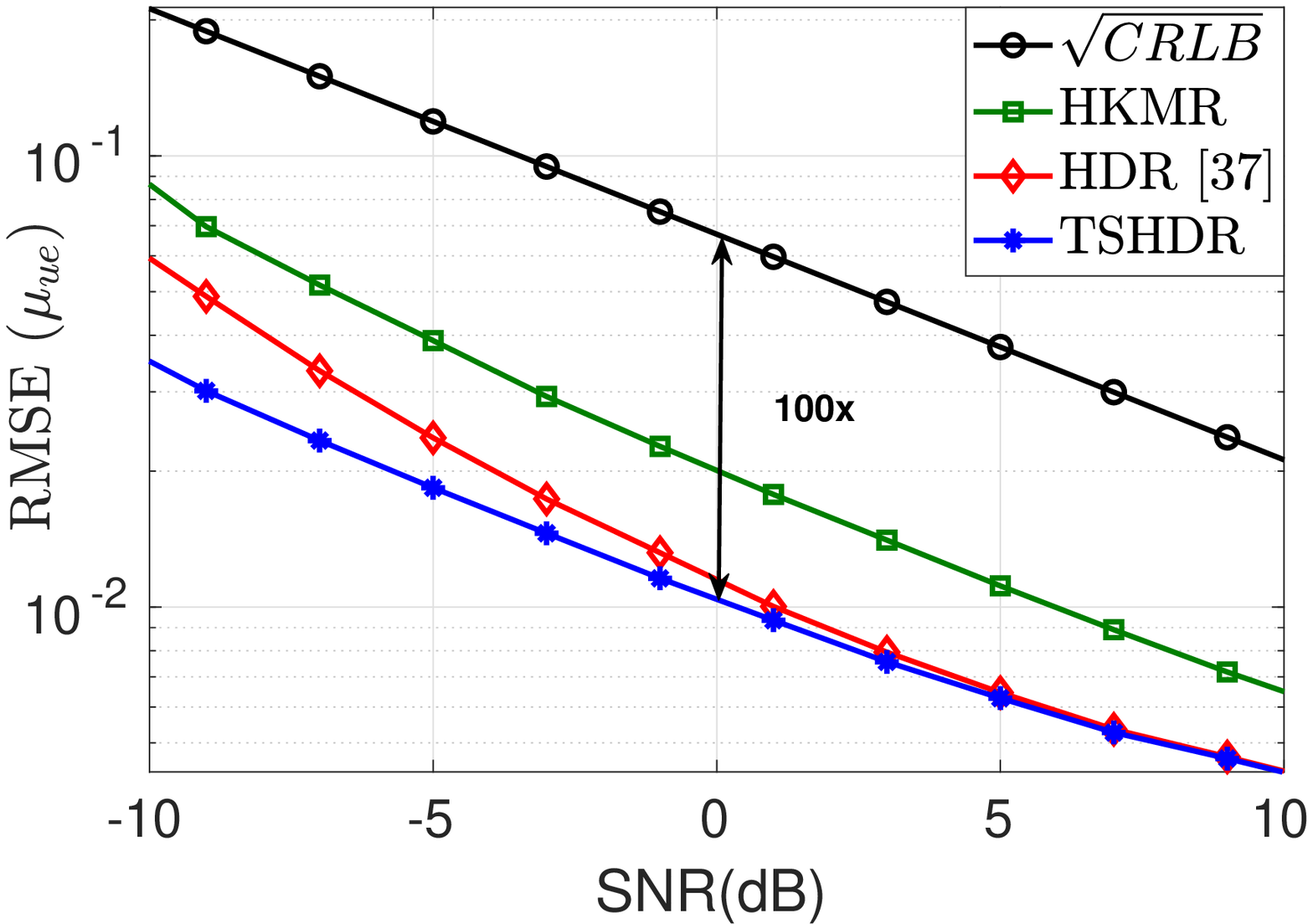}
		\caption{$\R{RMSE}$ of the $y$th spatial frequency $\mu_{\text{ue}}$}
		\label{fig:5_mu_ue}
	\end{subfigure}
	\begin{subfigure}[b]{0.45\textwidth}
		\centering
		\includegraphics[width=\textwidth]{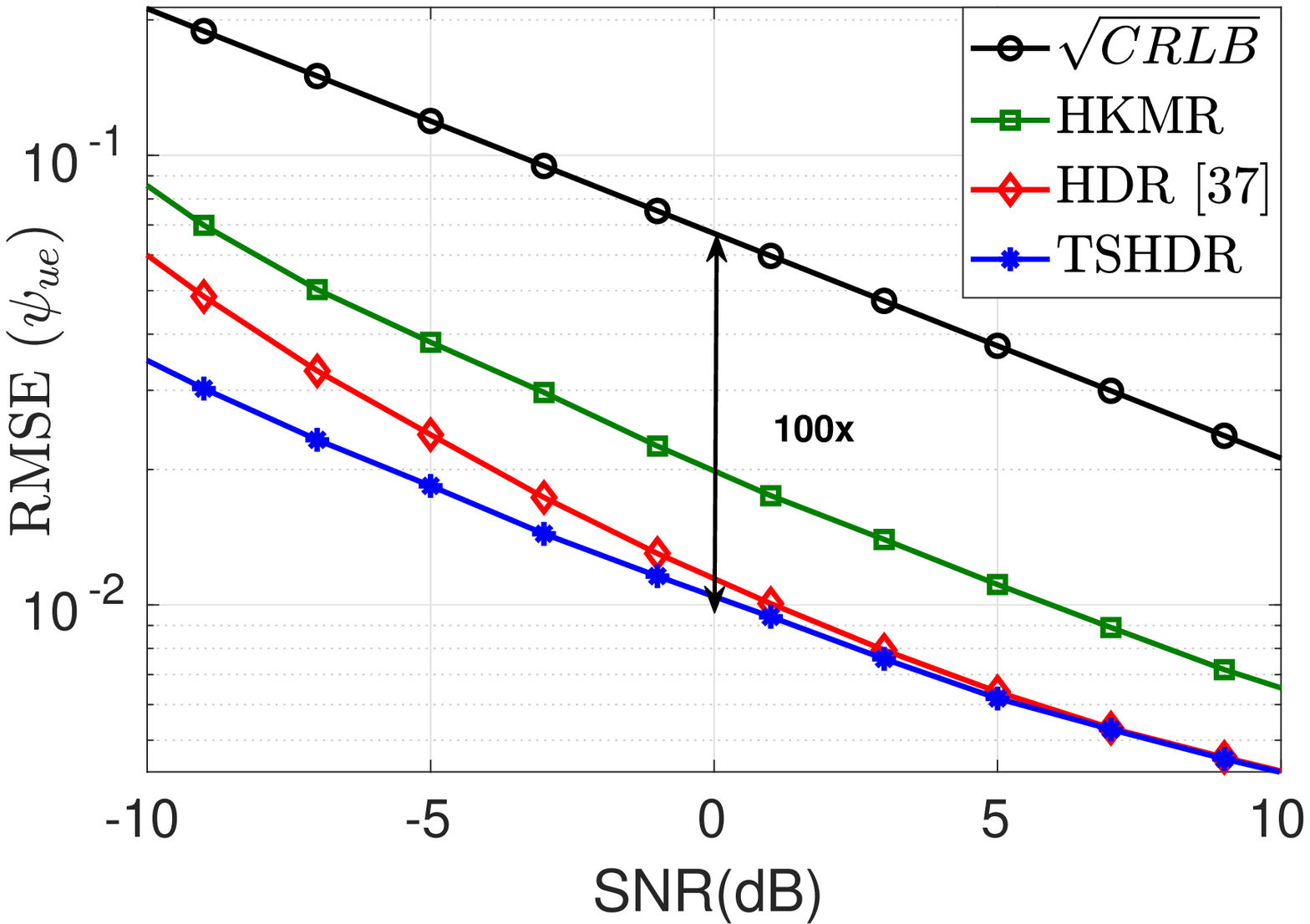}
		\caption{$\R{RMSE}$ of the $z$th spatial frequency $\psi_{\text{ue}}$}
		\label{fig:6_psi_ue}
	\end{subfigure}
	\caption{$\R{RMSE}$ for the estimated receive spatial frequencies associated with the UE assuming $M=16$, $Q=16$, and $N=16$.}
	\label{fig:ue}
\end{figure*}
The FIM $\bof{F}(\bof{\eta}_y)$ is given as
\begin{align}\label{FIM_y}
\bof{F}(\bof{\eta}_y) =%
\left[\begin{matrix}
	F_{\mu_\text{bs}\mu_\text{bs}} && F_{\mu_\text{bs}\mu_\text{ue}} && F_{\mu_\text{bs}\mu_\text{y}}\\
	F^\R{}_{\mu_\text{bs}\mu_\text{ue}} && F_{\mu_\text{ue} \mu_\text{ue}} && F_{\mu_\text{ue} \mu_\text{y}} \\
	F^\R{}_{\mu_\text{bs}\mu_\text{y}} && F^\R{}_{\mu_\text{ue} \mu_\text{y}} && F_{\mu_\text{y} \mu_\text{y}}
\end{matrix}\right], 
\end{align}
and
\begin{align}\label{FIM_z}
\bof{F}(\bof{\eta}_z) =%
\left[\begin{matrix}
	F_{\psi_\text{bs}\psi_\text{bs}} && F_{\psi_\text{bs}\psi_\text{ue}} && F_{\psi_\text{bs}\psi_\text{z}}\\
	F^\R{}_{\psi_\text{bs}\psi_\text{ue}} && F_{\psi_\text{ue} \psi_\text{ue}} && F_{\psi_\text{ue} \psi_\text{z}} \\
	F^\R{}_{\psi_\text{bs}\psi_\text{z}} && F^\R{}_{\psi_\text{ue} \psi_\text{z}} && F_{\psi_\text{z} \psi_\text{z}}
\end{matrix}\right]. 
\end{align}
The analytical expressions for the \al{derivation of the matrix} blocks of $\bof{F}(\bof{\eta}_y)$ and $\bof{F}(\bof{\eta}_z)$ are \al{provided} in Appendix \ref{Appendix_crlb}.
\section{Simulation Results}
\begin{figure}[!t]
\centerline{\includegraphics[width=0.7\linewidth]{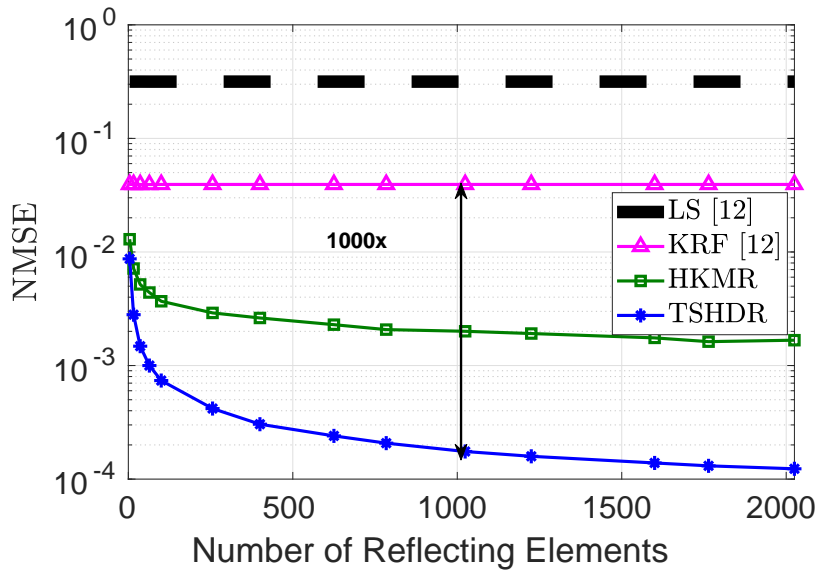}}
\caption{$\R{NMSE}$-based performance as a function of the number of IRS reflecting elements, assuming $\R{SNR} =\SI{5}{\decibel}$.}
\label{fig:a_nmsevsIRS}
\end{figure}

\begin{figure}[!t]
\centerline{\includegraphics[width=0.7\linewidth]{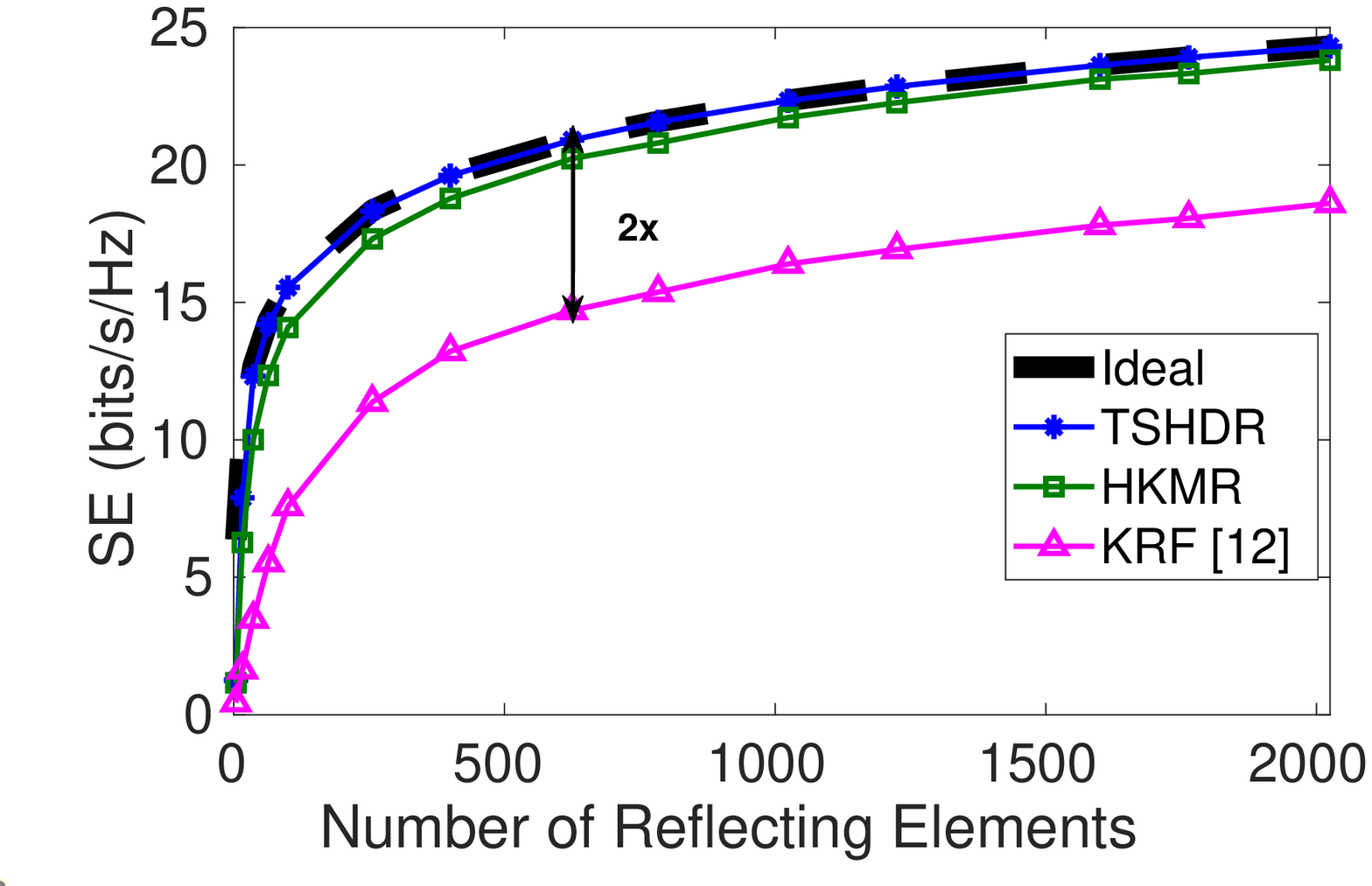}}
\caption{SE performance comparison by varying number of reflecting elements at $\R{SNR}=\SI{-17}{\decibel}$.}
\label{fig:a_sevsIRS}
\end{figure}

\begin{figure}[!t]
\centerline{\includegraphics[width=0.7\linewidth]{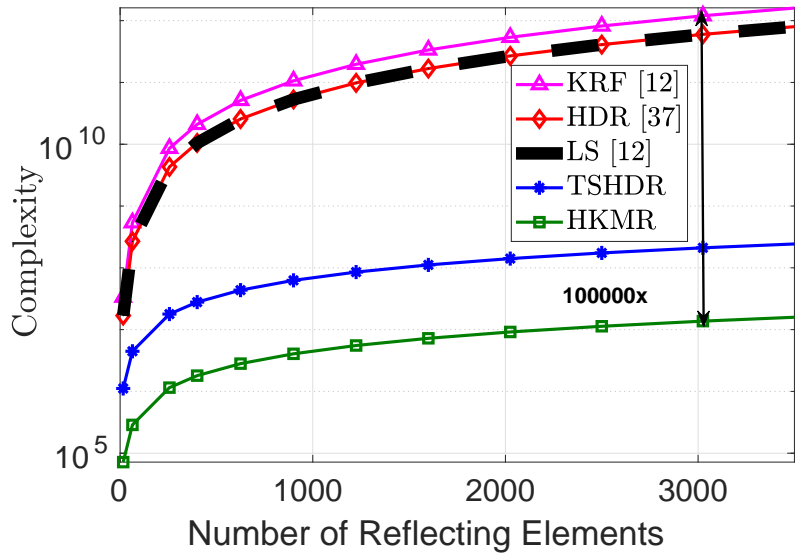}}
	\caption{\al{Computational complexity of HKMR and TSHDR, compared to the baseline LS method, the KRF method \cite{Gilderlan_2021} and the HDR method \cite{fazaleasim2023tensorbased}.}}
	\label{fig:a_complexity}
\end{figure}

In this section, we present the simulation results to evaluate the performance of our proposed estimators based on our design of pilots which consider the geometrical structure of the URA and use it to decouple the problem in horizontal and vertical domains. We assume the THz channels, where in most of the cases there is a dominant LOS path as given in \cite{Molisch_2021} and adopt a URA both at the BS and the UE. The AoD $\phi_{\text{bs}}$, $\phi_{\text{irs}_\text{D}}$ and AoA $\phi_{\text{irs}_\text{A}}$, $\phi_{\text{ue}}$ are uniformly generated assuming one sector of a cell as $\phi_{\text{bs}}, \phi_{\text{irs}_\text{D}}, \phi_{\text{irs}_\text{A}}, \phi_{\text{bs}} \sim \mathrm{U}\left(-60^\circ,+60^\circ\right)$. EoD $\theta_{\text{bs}}$, $\theta_{\text{irs}_\text{D}}$ and EoA $\theta_{\text{irs}_\text{A}}$, $\theta_{\text{ue}}$ are uniformly generated as $\theta_{\text{bs}}, \theta_{\text{irs}_\text{D}}, \theta_{\text{irs}_\text{A}}, \theta_{\text{bs}} \sim \mathrm{U}\left(90^\circ,130^\circ\right)$. Furthermore, assuming the number of antennas at the BS $M = 16$, where the number of antennas along the horizontal axis is $M_h=4$ and the number of antennas along the vertical axis is $M_v=4$. Similarly, the number of antennas at the UE is $Q= 16$ having $Q_h=4$ along the horizontal axis and $Q_v=4$ along the vertical axis. Finally, in all figures except Figs.~\ref{fig:a_nmsevsIRS}, \ref{fig:a_sevsIRS}, and \ref{fig:a_complexity} (which study the performance as a function of a varying number of IRS elements), the numbers of reflecting elements along the horizontal and vertical axes are $N_h=4$ and  $N_v=4$ respectively, implying $N=16$ reflecting elements in total. The total transmit power is assumed to be $P_T = \SI{1}{\watt}$ having transmit $\R{SNR} = P_T/\sigma_n^2$. 
The $\R{RMSE}$ for the channel parameters based on spatial frequencies is given as
\begin{align}
\R{RMSE}\left(x\right) = \sqrt{\mathbb{E}\left[ \left| x - \hat{x}\right |^2\right]}, \,\,\, x \in \left\{ \mu_{\text{bs}},\psi_{\text{bs}},\mu_{\text{ue}},\psi_{\text{ue}}, \mu_{\text{y}}, \psi_{\text{z}}  \right\}.\nonumber
\end{align}
We also evaluate the $\R{NMSE}$  associated with the global channel reconstructed from \al{its estimated horizontal and vertical components and parameters, defined as}
\begin{align}
\R{NMSE} = \mathbb{E} \left[ \frac{\left\|\bof{E} - \hat{\bof{E}} \right\|^2_2}{\left\| \bof{E} \right\|^2_2} \right],\nonumber
\end{align}
\al{where $\hat{\bof{E}}=\hat{\bof{H}}^\R{T} \diamond \hat{\bof{G}}$, with $\hat{\bof{H}}= \hat{\bof{H}}_y \otimes \hat{\bof{H}}_z$ and $\hat{\bof{G}}= \hat{\bof{G}}_y \otimes \hat{\bof{G}}_z$, which are built from the estimated angular parameters}.

\begin{figure*}
	\centering
	\begin{subfigure}[b]{0.45\textwidth}
		\centering
		\includegraphics[width=\textwidth]{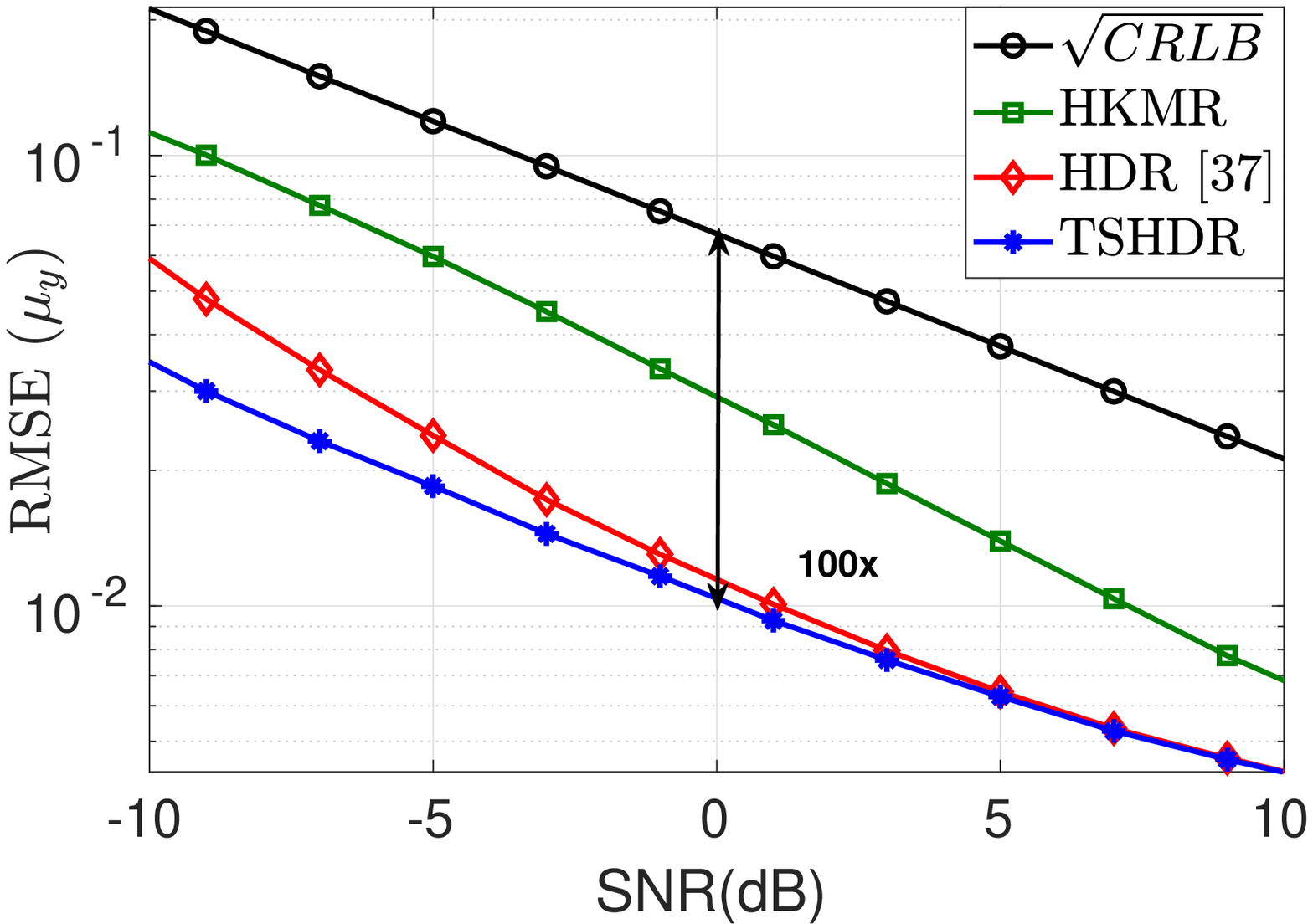}
		\caption{$\R{RMSE}$ of the $y$th combined spatial frequency $\mu_{\text{y}}$}
		\label{fig:7_mu_y}
	\end{subfigure}
	\begin{subfigure}[b]{0.45\textwidth}
		\centering
		\includegraphics[width=\textwidth]{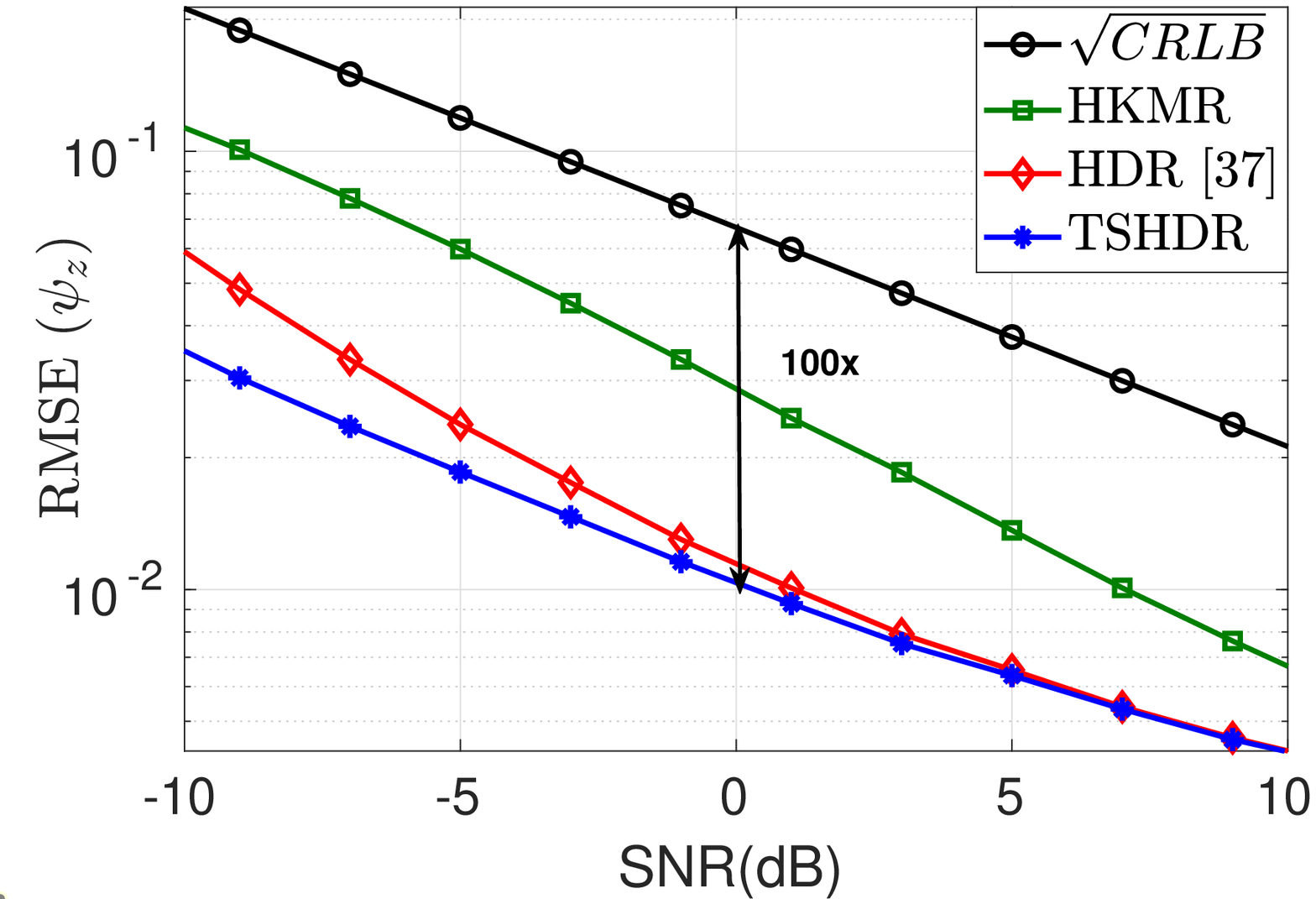}
		\caption{$\R{RMSE}$ of the $z$th combined spatial frequency $\psi_{\text{z}}$}
		\label{fig:8_psi_z}
	\end{subfigure}
	\caption{$\R{RMSE}$ of combined spatial frequencies of the IRS and the CRLB, using $N=16$ reflecting elements.}
	\label{fig:irs}
\end{figure*}
Figure~\ref{fig:bs} shows the performance of the HKMR and the TSHDR algorithms associated with the estimation of the transmit spatial frequencies \al{$\mu_{\text{bs}}$ and $\psi_{\text{bs}}$} based on the $\R{RMSE}$ metric. {\color{black} The two proposed estimators as well as the HDR algorithm \cite{fazaleasim2023tensorbased} outperform the $\text{CRLB}$. This is because all methods exploit the Kronecker structure of the involved channels \textit{via} rank-one approximation at the receiver, leading to noise rejection before the extraction of the channel parameters.} {\color{black} Note also that the TSHDR method outperforms both the HDR \cite{fazaleasim2023tensorbased} and HKMR methods. Such a gain comes from the more efficient noise rejection property of TSHDR provided by the Kronecker factorization of the $y$ and $z$ components of the large-dimensional combined channel matrix (step \ref{LS_algo_2} of Algorithm \ref{alg:TSHDR}). Although HKMR also has a Kronecker factorization procedure (step \ref{LSkron_alg_1} of Algorithm \ref{alg:HKMR}), the associated noise rejection is less efficient since this step is applied to the $y$ and $z$ components of the received pilot matrices, which have smaller dimensions. 
By its turn, the competing HDR method \cite{fazaleasim2023tensorbased} has no \textit{apriori} Kronecker factorization-based noise rejection stage, but involves factorization of sixth-order tensor, extracting all channel parameters at once, which explains its gain over HKMR.}
 \al{However, as we show later, HKMR has a significantly lower complexity compared to HDR \cite{fazaleasim2023tensorbased}, which is important for very large IRS panels. We can also note that for moderate/high SNRs,} the $\R{RMSE}$ of the HDR method converges \al{to that of the} TSHDR method. \al{Note that the TSHDR method has a gain of 100x compared to the CRLB for an SNR of \SI{0}{\decibel}. Moreover, for higher SNRs, all curves decrease linearly in the log domain with the same slope since noise rejection is irrelevant in the high-power regime.}

Figure~\ref{fig:ue} compares the proposed methods for the \fa{estimation of the receive spatial frequencies $\mu_{\text{ue}}$ and $\psi_{\text{ue}}$, in terms of the $\R{RMSE}$ performance metric. The results confirmed what has been discussed earlier, i.e., the two-stage TSHDR method outperforms the HKMR and HDR \cite{fazaleasim2023tensorbased} methods predominantly in the lower $\R{SNR}$ regime due to its two-stage noise rejection property. Otherwise stated, noise rejection is achieved from the LS Kronecker factorization in the first stage, and then in the second stage by means of the rank-one tensor approximation.} 
Overall, the proposed two methods as well as the HDR method, \cite{fazaleasim2023tensorbased} outperform the $\R{CRLB}$.
 For instance, with the parameter settings of Fig.~\ref{fig:6_psi_ue}, a gain of 100x in terms of $\R{RMSE}$ is observed at an SNR of \SI{0}{\decibel} for the TSHDR method, compared to the theoretical CRLB.

Figure~\ref{fig:irs} shows the $\R{RMSE}$ based performance comparison of HKMR and TSHDR methods for the estimation of combined $y$th spatial frequency $\mu_{\text{y}}$ and $z$th spatial frequency $\psi_{\text{z}}$. It is worth noting that HKMR performs a little poorly compared to the TSHDR and the HDR \cite{fazaleasim2023tensorbased} methods. {\color{black} This is due to some error propagation in the HKMR method since the BS/US and IRS channel parameters are estimated in two consecutive stages, as shown in steps \ref{alg_1_bf_comb}-\ref{alg_1_N} of Algorithm \ref{alg:HKMR}. On the other hand, TSDHR jointly estimates BS, UE, and IRS channel parameters (step 9 of Algorithm \ref{alg:TSHDR}).} 

\begin{figure}[!t]
\centerline{\includegraphics[width=0.7\linewidth]{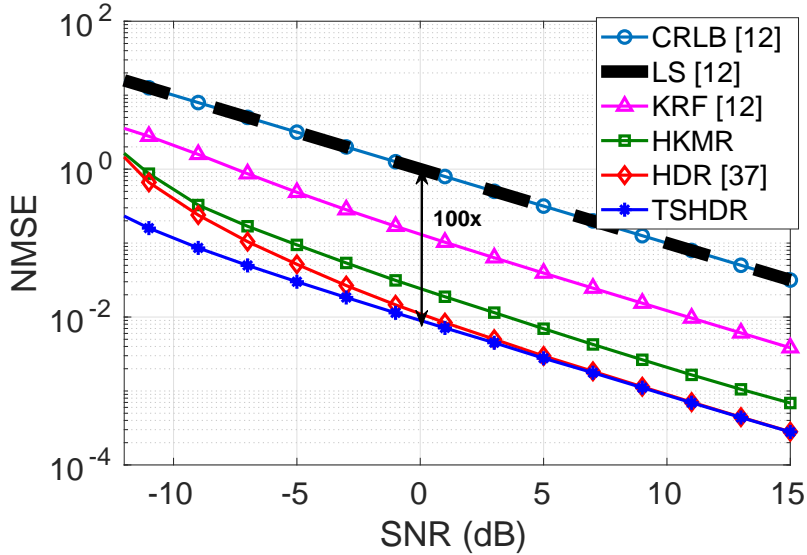}}
\caption{$\R{NMSE}$-based performance comparison of the whole reconstructed Khatri-Rao channel with HDR \cite{fazaleasim2023tensorbased}, LS, KRF methods and normalized CRLB \cite{Gilderlan_2021}.}
\label{fig:9_nmse}
\end{figure}

In Fig.~\ref{fig:9_nmse}, we evaluate the performance based on the $\R{NMSE}$ metric of the proposed HKMR and TSHDR methods with the competing methods, i.e., LS approach \cite{Gilderlan_2021}, KRF method \cite{Gilderlan_2021}, HDR method \cite{fazaleasim2023tensorbased}, and the normalized CRLB for the non-parametric channels is derived in \cite{Gilderlan_2021}. The Khatri-Rao channel given in \eqref{Khatri-Rao channel} is reconstructed using the estimated parameters. It is shown that the LS method satisfies the normalized CRLB while the rest of the algorithms outperform the normalized CRLB. Our proposed methods, i.e., HKMR, and TSHDR outperform the KRF method \cite{Gilderlan_2021}, in terms of NMSE. The reason is that the \fa{KRF method does not take into account the geometrical channel structure factorization, and consequently, does not exploit the resulting higher-order tensor structure of the problem.} While the HKMR method results in lower complexity compared to the HDR method \cite{fazaleasim2023tensorbased}, the HKMR shows a 
 \al{performance degradation} compared to TSHDR and HDR methods due to the propagation of estimation \al{errors from the first stage (concerned with the estimation of the BS and UE steering vectors) to the second stage (concerned with the estimation of IRS steering vectors).} 

Figure~\ref{fig:a_nmsevsIRS} shows the impact of  the number of IRS reflecting elements on the NMSE performance of the proposed and state-of-the-art methods. \fa{The results show that} the NMSE performance for the LS and the KRF methods \cite{Gilderlan_2021} are independent of the number of reflecting elements. {\color{black} The reason is that both methods do not exploit the Kronecker factorization structure of the cascaded Khatri-Rao channels in the estimation process. On the other hand, the HKMR and TSHDR methods exploit the two-dimensional $y$ and $z$ decomposition of the IRS phase shift matrices, which leads to $y$th and $z$th rank-one tensor factorization problems that result in more efficient noise rejection, ultimately improving the overall estimation accuracy.
For instance, assuming $N$=1000 IRS elements, the performance gain of TSHDR over KRF \cite{Gilderlan_2021} is approximately 1000 times.
Note that under optimal IRS training design, the NMSE of LS and KRF methods do not depend on the number of IRS elements. The HKMR and HTSHDR methods benefit from the increase in the number of IRS elements to improve the NMSE performance, especially in the range of up to 500 elements. Indeed, the rank-one matrix approximation step (step 12 in Algorithm \ref{alg:HKMR}) and the rank-one tensor approximation step (step 9 in Algorithm \ref{alg:TSHDR}) lead to more accurate estimates of the channel's steering vectors as $N_y$ and $N_z$ as increased.} 

\begin{figure}[!t]
\centerline{\includegraphics[width=0.7\linewidth]{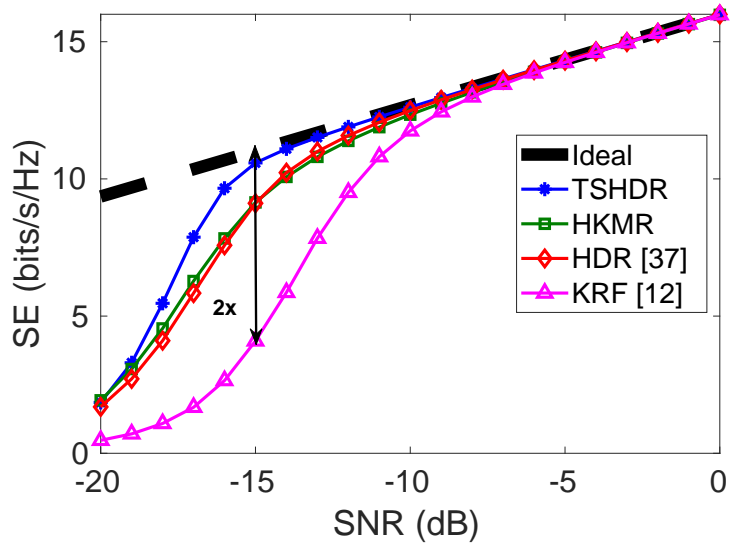}}
\caption{SE performance of the proposed algorithms under actual channel estimation.}
\label{fig:10_se}
\end{figure}

\al{In Figures \ref{fig:10_se} and \ref{fig:a_sevsIRS},} the performance of proposed algorithms is \al{assessed} using SE as the performance metric. The design of both the combiner and the precoder is based on the reconstructed estimated channels. \al{As a reference for comparison, we also plot the ideal case where the precoder and combiner are derived from the true channel. We can note that the} TSHDR method yields the best SE performance, especially in the low SNR regime. At moderate/high SNRs, the proposed methods perform satisfactorily, approaching the sum rate of the perfect channel knowledge case. By contrast, the KRF method still leads to a sum rate drop even at moderate SNRs.

Figure ~\ref{fig:a_sevsIRS} shows the impact of the number of reflecting elements on sum rate performance. \al{We can see} that increasing the number of reflecting elements yields an improvement in the sum rate due to the higher spatial resolution. Furthermore, it can be noted that the TSHDR method outperforms the KRF one. For instance, considering the parameter settings of Fig.~\ref{fig:a_sevsIRS} and an SNR of \SI{-17}{\decibel}, the sum rate provided by the TSHDR method is twice that obtained with the KRF method.

Figure~\ref{fig:a_complexity} shows a complexity comparison of the two proposed algorithms, i.e., HKMR and TSHDR. We also plot the complexity of the reference channel estimation methods, namely the LS \cite{Gilderlan_2021}, the KRF \cite{Gilderlan_2021} methods, and the HDR method \cite{fazaleasim2023tensorbased}. Due to the decoupling of the global problem into horizontal and vertical domains sub-problems, the computational complexity of the proposed algorithms is much lower than that of the KRF, LS, and HDR methods. Moreover, the HKMR method offers the lowest complexity among its competitors. For instance, assuming 3000 reflecting elements, the HKMR method is $10^5$ times less complex than the KRF method \cite{Gilderlan_2021}, which is a remarkable result. We can also see that the complexity of the HDR method \cite{fazaleasim2023tensorbased} is similar to that of the LS method. On the other hand, the TSHDR method is slightly more complex than the HKMR method. However, it is worth noting that such a small extra complexity comes with a significant performance improvement, as can be seen in our previous numerical results (Figs.~\ref{fig:bs} to \ref{fig:a_sevsIRS}). {\color{black} By its turn, the TSHDR method shows similar performance as the competing HDR method in moderate to high SNRs but offers a significant complexity reduction. These different results corroborate the benefits of the proposed two-dimensional structured channel estimation methods while highlighting their tradeoffs and scenarios of interest.}

\section{Conclusions}
This paper proposed two algorithms for two-dimensional channel parameter estimation in IRS-assisted MIMO communications by exploiting the geometry of the involved channels as well as their algebraic structures. We showed that the global parameter estimation problem can be decoupled into horizontal and vertical domains by designing an independent pilot matrix along each domain. Such a decoupling allows the formulation of the structured channel estimation as a Kronecker factorization problem involving the horizontal and vertical components. By exploiting the algebraic structure of the received two-dimensional pilots, we recast this problem as separate rank-one tensor approximation problems. Our proposed HKMR and TSHDR methods exploit the two-dimensional pilot structure along each domain and split the single (bigger) channel estimation problem into two smaller sub-problems. Both methods offer significantly more accurate channel parameter estimates and higher spectral efficiency than the competing state-of-the-art LS, the KRF, and HDR methods with lower complexity, thus being attractive solutions to structured channel parameter estimation, especially for large IRS panels. \al{Perspectives of this work include the generalization of the proposed methods to advanced IRS architectures such as hybrid IRS \cite{Zhang_2023} and simultaneous transmitting and reflecting (STAR) IRS \cite{Mu_2022} Further work shall include time-varying channels to capture UE mobility as extensions to the multi-user case.}

\appendices
\section{Entries of the FIM $\B{F}(\bof{\eta}_y)$ and FIM $\B{F}(\bof{\eta}_z)$} \label{Appendix_crlb}
The entries of the block matrices of the two FIMs defined in \eqref{FIM_y} and \eqref{FIM_z} are derived as follows
\begin{align}
F_{\mu_{\text{bs}}\mu_{\text{bs}}}=  \frac{2}{\sigma_{n}^2}\R{Re}\left(\R{tr}\left\{\frac{\partial \left(\left(\bof{a}_y(\mu_{\text{bs}}) \otimes  \bof{q}_y(\mu_{\text{ue}})\right) \bof{n}_y^\R{T}(\mu_{\text{y}})\right)^\R{H}}{\partial\mu_{\text{bs}}}
\frac{\partial \left(\left(\bof{a}_y(\mu_{\text{bs}}) \otimes  \bof{q}_y(\mu_{\text{ue}})\right) \bof{n}_y^\R{T}(\mu_{\text{y}})\right)}{\partial \mu_{\text{bs}}}\right\}\right),\nonumber
\end{align}
\begin{align}
F_{\mu_{\text{bs}}\mu_{\text{ue}}}=  \frac{2}{\sigma_{n}^2}\R{Re}\left(\R{tr}\left\{\frac{\partial \left(\left(\bof{a}_y(\mu_{\text{bs}}) \otimes  \bof{q}_y(\mu_{\text{ue}})\right) \bof{n}_y^\R{T}(\mu_{\text{y}})\right)^\R{H}}{\partial\mu_{\text{bs}}}
 \frac{\partial \left(\left(\bof{a}_y(\mu_{\text{bs}}) \otimes  \bof{q}_y(\mu_{\text{ue}})\right) \bof{n}_y^\R{T}(\mu_{\text{y}})\right)}{\partial \mu_{\text{ue}}}\right\}\right),\nonumber
\end{align}
\begin{align}
F_{\mu_{\text{bs}}\mu_{\text{y}}}=  \frac{2}{\sigma_{n}^2}\R{Re}\left(\R{tr}\left\{\frac{\partial \left(\left(\bof{a}_y(\mu_{\text{bs}}) \otimes  \bof{q}_y(\mu_{\text{ue}})\right) \bof{n}_y^\R{T}(\mu_{\text{y}})\right)^\R{H}}{\partial\mu_{\text{bs}}}
 \frac{\partial \left(\left(\bof{a}_y(\mu_{\text{bs}}) \otimes  \bof{q}_y(\mu_{\text{ue}})\right) \bof{n}_y^\R{T}(\mu_{\text{y}})\right)}{\partial \mu_{\text{y}}}\right\}\right),\nonumber
\end{align}
\begin{align}
F_{\mu_{\text{ue}}\mu_{\text{ue}}}=  \frac{2}{\sigma_{n}^2}\R{Re}\left(\R{tr}\left\{\frac{\partial \left(\left(\bof{a}_y(\mu_{\text{bs}}) \otimes  \bof{q}_y(\mu_{\text{ue}})\right) \bof{n}_y^\R{T}(\mu_{\text{y}})\right)^\R{H}}{\partial\mu_{\text{ue}}}
 \frac{\partial \left(\left(\bof{a}_y(\mu_{\text{bs}}) \otimes  \bof{q}_y(\mu_{\text{ue}})\right) \bof{n}_y^\R{T}(\mu_{\text{y}})\right)}{\partial \mu_{\text{ue}}}\right\}\right),\nonumber
\end{align}
\begin{align}
F_{\mu_{\text{ue}}\mu_{\text{y}}}=  \frac{2}{\sigma_{n}^2}\R{Re}\left(\R{tr}\left\{\frac{\partial \left(\left(\bof{a}_y(\mu_{\text{bs}}) \otimes  \bof{q}_y(\mu_{\text{ue}})\right) \bof{n}_y^\R{T}(\mu_{\text{y}})\right)^\R{H}}{\partial\mu_{\text{ue}}}
 \frac{\partial \left(\left(\bof{a}_y(\mu_{\text{bs}}) \otimes  \bof{q}_y(\mu_{\text{ue}})\right) \bof{n}_y^\R{T}(\mu_{\text{y}})\right)}{\partial \mu_{\text{y}}}\right\}\right),\nonumber
\end{align}
\begin{align}
F_{\mu_{\text{y}}\mu_{\text{y}}}=  \frac{2}{\sigma_{n}^2}\R{Re}\left(\R{tr}\left\{\frac{\partial \left(\left(\bof{a}_y(\mu_{\text{bs}}) \otimes  \bof{q}_y(\mu_{\text{ue}})\right) \bof{n}_y^\R{T}(\mu_{\text{y}})\right)^\R{H}}{\partial\mu_{\text{y}}}
 \frac{\partial \left(\left(\bof{a}_y(\mu_{\text{bs}}) \otimes  \bof{q}_y(\mu_{\text{ue}})\right) \bof{n}_y^\R{T}(\mu_{\text{y}})\right)}{\partial \mu_{\text{y}}}\right\}\right),\nonumber
\end{align}

Similarly, the entries of the block matrices of the FIM \eqref{FIM_z} are derived as,
\begin{align}
F_{\psi_{\text{bs}}\psi_{\text{bs}}}=  \frac{2}{\sigma_{n}^2}\R{Re}\left(\R{tr}\left\{\frac{\partial \left(\left(\bof{a}_z(\psi_{\text{bs}}) \otimes  \bof{q}_z(\psi_{\text{ue}})\right) \bof{n}_z^\R{T}(\psi_{\text{z}})\right)^\R{H}}{\partial\psi_{\text{bs}}}
 \frac{\partial \left(\left(\bof{a}_z(\psi_{\text{bs}}) \otimes  \bof{q}_z(\psi_{\text{ue}})\right) \bof{n}_z^\R{T}(\psi_{\text{z}})\right)}{\partial \psi_{\text{bs}}}\right\}\right),\nonumber
\end{align}
\begin{align}
F_{\psi_{\text{bs}}\psi_{\text{ue}}}=  \frac{2}{\sigma_{n}^2}\R{Re}\left(\R{tr}\left\{\frac{\partial \left(\left(\bof{a}_z(\psi_{\text{bs}}) \otimes  \bof{q}_z(\psi_{\text{ue}})\right) \bof{n}_z^\R{T}(\psi_{\text{z}})\right)^\R{H}}{\partial\psi_{\text{bs}}}
 \frac{\partial \left(\left(\bof{a}_z(\psi_{\text{bs}}) \otimes  \bof{q}_z(\psi_{\text{ue}})\right) \bof{n}_z^\R{T}(\psi_{\text{z}})\right)}{\partial \psi_{\text{ue}}}\right\}\right),\nonumber
\end{align}
\begin{align}
F_{\psi_{\text{bs}}\psi_{\text{z}}}=  \frac{2}{\sigma_{n}^2}\R{Re}\left(\R{tr}\left\{\frac{\partial \left(\left(\bof{a}_z(\psi_{\text{bs}}) \otimes  \bof{q}_z(\psi_{\text{ue}})\right) \bof{n}_z^\R{T}(\psi_{\text{z}})\right)^\R{H}}{\partial\psi_{\text{bs}}}
 \frac{\partial \left(\left(\bof{a}_z(\psi_{\text{bs}}) \otimes  \bof{q}_z(\psi_{\text{ue}})\right) \bof{n}_z^\R{T}(\psi_{\text{z}})\right)}{\partial \psi_{\text{z}}}\right\}\right),\nonumber
\end{align}
\begin{align}
F_{\psi_{\text{ue}}\psi_{\text{ue}}}=  \frac{2}{\sigma_{n}^2}\R{Re}\left(\R{tr}\left\{\frac{\partial \left(\left(\bof{a}_z(\psi_{\text{bs}}) \otimes  \bof{q}_z(\psi_{\text{ue}})\right) \bof{n}_z^\R{T}(\psi_{\text{z}})\right)^\R{H}}{\partial\psi_{\text{ue}}}
 \frac{\partial \left(\left(\bof{a}_z(\psi_{\text{bs}}) \otimes  \bof{q}_z(\psi_{\text{ue}})\right) \bof{n}_z^\R{T}(\psi_{\text{z}})\right)}{\partial \psi_{\text{ue}}}\right\}\right),\nonumber
\end{align}
\begin{align}
F_{\psi_{\text{ue}}\psi_{\text{z}}}=  \frac{2}{\sigma_{n}^2}\R{Re}\left(\R{tr}\left\{\frac{\partial \left(\left(\bof{a}_z(\psi_{\text{bs}}) \otimes  \bof{q}_z(\psi_{\text{ue}})\right) \bof{n}_z^\R{T}(\psi_{\text{z}})\right)^\R{H}}{\partial\psi_{\text{ue}}}
 \frac{\partial \left(\left(\bof{a}_z(\psi_{\text{bs}}) \otimes  \bof{q}_z(\psi_{\text{ue}})\right) \bof{n}_z^\R{T}(\psi_{\text{z}})\right)}{\partial \psi_{\text{z}}}\right\}\right),\nonumber
\end{align}
\begin{align}
F_{\psi_{\text{z}}\psi_{\text{z}}}=  \frac{2}{\sigma_{n}^2}\R{Re}\left(\R{tr}\left\{\frac{\partial \left(\left(\bof{a}_z(\psi_{\text{bs}}) \otimes  \bof{q}_z(\psi_{\text{ue}})\right) \bof{n}_z^\R{T}(\psi_{\text{z}})\right)^\R{H}}{\partial\psi_{\text{z}}}
 \frac{\partial \left(\left(\bof{a}_z(\psi_{\text{bs}}) \otimes  \bof{q}_z(\psi_{\text{ue}})\right) \bof{n}_z^\R{T}(\psi_{\text{z}})\right)}{\partial \psi_{\text{z}}}\right\}\right).\nonumber
\end{align}
The partial derivative of 
\begin{align}
\frac{\partial \left(\left(\bof{a}_y(\mu_{\text{bs}}) \otimes  \bof{q}_y(\mu_{\text{ue}})\right) \bof{n}_y^\R{T}(\mu_{\text{y}})\right)^\R{H}}{\partial \mu_{\text{bs}}}  = 
  \frac{\partial \bof{n}_y^\ast (\mu_{\text{y}}) \left( \bof{a}_y^\R{H}(\mu_{\text{bs}}) \otimes \bof{q}_y^\R{H}(\mu_{\text{ue}}) \right)}{\partial \mu_{\text{bs}}} = 
 \bof{n}_y^\ast (\mu_{\text{y}}) \left( \bof{a}_y^{\prime\R{H}}(\mu_{\text{bs}}) \otimes \bof{q}_y^\R{H}(\mu_{\text{ue}}) \right),\nonumber
\end{align}
where
\begin{align}
\bof{a}_y^\prime(\mu_{\text{bs}}) = 	\frac{\partial \bof{a}_y(\mu_{\text{bs}})}{\partial \mu_{\text{bs}}} = \left[0, (-j)e^{(-j\mu_{\text{bs}})},\dots ,(-j(M_y-1))e^{-j(
	M_y-1)\mu_{\text{bs}}}\right]^\R{T}.\nonumber
\end{align}

\bibliographystyle{IEEEtran}
\bibliography{IEEEabrv,globcombib}

\end{document}